\input harvmac.tex
\input epsf
\noblackbox

\newcount\figno

\figno=0
\def\fig#1#2#3{
\par\begingroup\parindent=0pt\leftskip=1cm\rightskip=1cm\parindent=0pt
\baselineskip=11pt \global\advance\figno by 1 \midinsert
\epsfxsize=#3 \centerline{\epsfbox{#2}} \vskip 12pt
\centerline{{\bf Figure \the\figno :}{\it ~~ #1}}\par
\endinsert\endgroup\par}
\def\figlabel#1{\xdef#1{\the\figno}}


\def\half{{1\over 2}}

\def\inbar{\vrule height1.5ex width.4pt depth0pt}
\def\IC{\relax\,\hbox{$\inbar\kern-.3em{\rm C}$}}
\def\IR{\relax{\rm I\kern-.18em R}}
\def\IN{\relax{\rm I\kern-.18em N}}
\def\IP{\relax{\rm I\kern-.18em P}}
\def\ZZ{\bb{Z}}
\def\narrowplus{\kern -.04truein + \kern -.03truein}
\def\narrowminus{- \kern -.04truein}
\def\narrowminussub{\kern -.02truein - \kern -.01truein}

\def\Im{{\rm Im}}

\def\P{I\hskip -2pt P}

\font\mbm = msbm10
\def\bb#1{\hbox{\mbm #1}}

\lref\russo{R. Russo and S. Sciuto, Nucl.\ Phys.\ B 669 (2003) 207,
hep-th/0306129; Fortsch. Phys. 52 (2004) 678, hep-th/0312205.}
\lref\atick{J.J. Atick, L.J. Dixon and P.A. Griffin,
Nucl.\ Phys.\ B 298 (1988) 1.}
\lref\fay{John D. Fay, {\it Theta Functions on Riemann Surfaces}, in {\it
Lecture
Notes in Mathematics} (Springer-Verlag, 1973).}
\lref\mumford{David Mumford, {\it Tata Lectures on Theta II} (Birkh\"auser,
1993).}
\lref\ver{E. Verlinde and H. Verlinde, Nucl.\ Phys.\ B 288 (1987) 357.}
\lref\dhok{E. D'Hoker and D.H. Phong,
Nucl.\ Phys.\ B 639 (2002) 129, hep-th/0111040.}
\lref\narain{K.~S.~Narain, M.~H.~Sarmadi and C.~Vafa,
Nucl.\ Phys.\ B 356 (1991) 163.}
\lref\morozov{A.~A.~Belavin, V.~Knizhnik, A.~Morozov and A.~Perelomov,
JETP Lett.\  43 (1986) 411
[Phys.\ Lett.\ B 177 (1986) 324].}
\lref\moore{G. Moore, Phys.\ Lett.\ B 176 (1986) 369.}
\lref\SS{J.~Scherk and J.~H.~Schwarz,
Phys.\ Lett.\ B 82 (1979) 60.}
\lref\rt{R.~Rohm,
Nucl.\ Phys.\ B 237 (1984) 553; H. Itoyama and T.R. Taylor, Phys.\
Lett.\ B 186 (1987) 129.} \lref\kpor{C.~Kounnas and M.~Porrati,
Nucl.\ Phys.\ B 310 (1988) 355.}
  \lref\SSII{
S.~Ferrara, C.~Kounnas, M.~Porrati and F.~Zwirner,
Nucl.\ Phys.\ B 318 (1989) 75;
C.~Kounnas and B.~Rostand,
Nucl.\ Phys.\ B 341 (1990) 641;
I.~Antoniadis,
Phys.\ Lett.\ B 246 (1990)~377.}
\lref\SSI{I.~Antoniadis, E.~Dudas and A.~Sagnotti,
Nucl.\ Phys.\ B 544 (1999) 469, hep-th/9807011.}
\lref\ABLM{
I.~Antoniadis, K.~Benakli, A.~Laugier and T.~Maillard,
Nucl.\ Phys.\ B 662 (2003) 40, hep-ph/0211409.}
\lref\invo{
S.~K.~Blau, S.~Carlip, M.~Clements, S.~Della Pietra and V.~Della Pietra,
Nucl.\ Phys.\ B 301 (1988) 285;
M.~Bianchi and A.~Sagnotti,
Phys.\ Lett.\ B 211 (1988) 407.}
\lref\modular{
M.~Bianchi and A.~Sagnotti,
Phys.\ Lett.\ B 231 (1989) 389.}
\lref\ant{
I.~Antoniadis, K.~S.~Narain and T.R.~Taylor, in preparation.}
\lref\Oog{
H.~Ooguri and C.~Vafa, hep-th/0302109.}
\lref\bcov{M. Bershadsky, S. Cecotti, H. Ooguri and C. Vafa, Commun.
Math Phys. 165 (1994) 311, hep-th/9309140.}
\lref\agnt{
I.~Antoniadis, E.~Gava, K.~S.~Narain and T.R.~Taylor,
Nucl.\ Phys.\ B 413 (1994) 162, hep-th/9307158.}
\lref\aq{
I.~Antoniadis and M.~Quiros,
Nucl.\ Phys.\ B {505} (1997) 109, hep-th/9705037;
R.~Rattazzi, C.~A.~Scrucca and A.~Strumia,
Nucl.\ Phys.\ B {674} (2003) 171, hep-th/0305184.}
\lref\rs{
L.~Randall and R.~Sundrum,
Nucl.\ Phys.\ B {557} (1999) 79, hep-th/9810155; G. Giudice, M.A. Luty,
H. Murayama and R. Rattazzi, JHEP 9812 (1998) 027, hep-ph/9810442.}
\lref\dfms{
L.~Dixon, D.~Friedan, M.~Martinec and S.~Shenker,
Nucl.\ Phys.\ B {282} (1987) 13.}
\lref\bsb{
S.~Sugimoto,
Prog.\ Theor.\ Phys.\ 102 (1999) 685, hep-th/9905159;
I.~Antoniadis, E.~Dudas and A.~Sagnotti,
Phys.\ Lett.\ B {464} (1999) 38, hep-th/9908023;
G.~Aldazabal and A.~M.~Uranga,
JHEP {9910} (1999) 024, hep-th/9908072.} \lref\nl{ E.~Dudas and
J.~Mourad, Phys.\ Lett.\ B {514} (2001) 173, hep-th/0012071;
G.~Pradisi and F.~Riccioni,
Nucl.\ Phys.\ B {615} (2001) 33, hep-th/0107090.}

\Title{\vbox{
      \hbox{CERN--PH--TH/2004-060}
     \hbox{hep-th/0403293}}}
{\vbox{\centerline{Topological Masses From Broken Supersymmetry}}}
\centerline{Ignatios Antoniadis{$^{1,\dagger}$} and Tomasz
R.\ Taylor{$^{2}$ }}
\bigskip\medskip
\centerline{$^1${\it CERN Theory Division, CH-1211
Geneva 23, Switzerland}}
\centerline{$^2${\it Department of Physics, Northeastern University, Boston, MA
02115, U.S.A.}}
\bigskip
\bigskip

\centerline{\bf Abstract} \noindent We develop a formalism for
computing one-loop gravitational corrections to the effective
action of D-branes. In particular, we study bulk to brane
mediation of supersymmetry breaking in models where supersymmetry
is broken at the tree-level in the closed string sector (bulk) by
Scherk-Schwarz boundary conditions, while it is realized on a
collection of D-branes in a linear or non-linear way. We compute
the gravitational corrections to the fermion masses $m_{1/2}$
(gauginos or goldstino) induced from the exchange of closed
strings, which are non-vanishing for world-sheets with Euler
characteristic ${-}1$ (``genus 3/2'') due to a string diagram with
one handle and one hole. We show that the corrections have a
topological origin and that in general, for a small gravitino
mass, the induced mass behaves as $m_{1/2}\propto g^4 m_{3/2}$,
with $g$ the gauge coupling. In generic orbifold compactifications
however, this leading term vanishes as a consequence of cancellations
caused by discrete symmetries, and the remainder is exponentially
suppressed by a factor of $\exp(-1/\alpha'm^2_{3/2})$.

\vfill\hrule\vskip 1mm\noindent
{$^{\dagger}$}\ninerm On leave from CPHT Ecole Polytechnique
(UMR du CNRS 7644) F-91128, Palaiseau.\hfil\break
\Date{}
\newsec{Introduction}

There are many important reasons for studying gravitational
corrections to the effective actions describing D-brane
excitations. One of the phenomenologically-oriented goals
presenting quite a challenging theoretical problem is to
understand how supersymmetry breaking in the bulk can possibly be
communicated to the gauge theory of supersymmetric branes.
The world-sheet configuration that mediates such a supersymmetry
breaking is a bordered Riemann surface with one hole and one
handle (or two crosscaps in the presence of orientifolds) and has
Euler characteristic ${-}1$: we call it a ``genus 3/2'' surface.
In this work, we develop a formalism for such computations and
study the (Majorana) fermion masses on branes generated from
gravitational corrections in type II string models with supersymmetry
breaking realized via Scherk-Schwarz boundary conditions
\refs{\SS,\rt,\kpor,\SSII,\SSI}. The gravitino mass is then
proportional to the compactification scale $1/R$, while
supersymmetry remains unbroken locally on the world-volume of
branes transverse to the Scherk-Schwarz (SS)
direction~\refs{\SSI}.

Actually, in type I theory, there is also a pair of orientifold
$({\cal O})$ and anti-orientifold $(\bar{\cal O})$ planes formed,
respectively, at the two endpoints of the compactification interval,
requiring in general the presence of branes and anti-branes for canceling
the Ramond-Ramond (RR) charge. One can then construct two types of
D-brane configurations~\ABLM: D${\cal O}$ or ${\bar{\rm D}}\bar{\cal O}$
preserving linearly half of the bulk supersymmetry on the brane
world-volume~\refs{\SSI}, and ${\bar{\rm D}}{\cal O}$ or D$\bar{\cal O}$
with non supersymmetric spectra but realizing non-linear
supersymmetry~\refs{\bsb,\nl}. Our study of gravitational corrections to
Majorana fermion masses applies both to gauginos of the supersymmetric
configurations, as well as to the goldstino of the non-supersymmetric
models.

Since the mediation of supersymmetry breaking from the bulk is a
local phenomenon, the possible existence of distinct branes and
orientifolds far away in the bulk can only lead to a secondary
subleading effect. Thus, for simplicity and without loss of
generality, we will restrict our analysis to the oriented string
case, the generalization in the presence of orientifolds being
straightforward. Moreover, for the same reason, we will ignore the
possible existence of distinct anti-branes which are required by
RR charge conservation in the compact case. Note that the brane --
antibrane system is not affected by the SS deformation. In any
case, their presence can be avoided if some other direction of the
transverse space is non-compact, without affecting our results.

Now consider a fermion mass term generated by the two-point
function at zero momentum of two (tree-level) massless gauginos
(or goldstinos) of the same (four-dimensional) chirality. The
generic oriented string diagram that contributes to this amplitude
has $g$ handles and $h$ holes corresponding to D-brane boundaries.
Obviously, one must have $h\ge 1$ in order to insert the gaugino
vertices. The power of the string couplings is determined by the
Euler characteristic $\chi=2-2g-h$. A simple inspection of the
internal $N=2$ world-sheet superconformal charge, of which each
gaugino vertex carries 3/2 units, shows that one needs at least
three world-sheet supercurrent insertions to ensure charge
conservation. This implies that independently of the source of
bulk supersymmetry breaking, the lowest order diagram that can
give non-vanishing contribution has $\chi={-}1$ and ``genus''
$g+h/2= 3/2$. In the oriented string case, there are two such
diagrams: one with three boundaries $(g=0,h=3)$ and one with a
handle and a boundary $(g=1,h=1)$. In the framework we described
above, all boundaries are of the same type and thus the first
diagram is supersymmetric and can be ignored. We are therefore
left with the $(g=1,h=1)$ surface that contains the information
about gravitational interactions. This world-sheet configuration
will be studied in this work in great detail.

The main result of our analysis is that the  gaugino mass is
determined by an amplitude closely related to the topological partition
function ${\cal F}_2$ \refs{\bcov,\agnt}. Furthermore, for smooth
compactifications with at most $N=2$ supersymmetry on the branes,
we obtain, in the limit of low-energy supersymmetry breaking, a
fermion mass proportional to the gravitino mass $m_{3/2}$.  On the
other hand, within the effective field theory, the one loop
gravitational correction leads obviously to two inverse powers of
the Planck mass which can be canceled only if the momentum
integral is quadratically divergent. Thus the leading contribution
computes precisely the coefficient of this divergence, cutoff by
the string scale. However, in the case
of orbifolds, this term vanishes as a result of cancellations
caused by discrete symmetries, and the remainder is suppressed
exponentially as $\exp (-1/\alpha' m_{3/2}^2)$.

The paper is organized as follows. In Section 2, we describe the
moduli space of the relevant genus 3/2 string diagram $(g=1,h=1)$,
obtained by an appropriate involution of the genus 2 $(g=2,h=0)$
surface. We also obtain the left-over modular group and the
corresponding fundamental domain of integration. In Section 3, we
derive the partition function modified by the SS deformation and
study the degeneration regions associated to the large radius
limit. In Section 4, we compute the amplitude that determines the
gaugino mass. We point out that it is closely related to the
amplitude corresponding to the topological ${\cal F}_2$ term in
type II theory. To simplify our discussion, we assume that the
action of the SS twist is limited to the two-torus part of a
$T^2\times K_3$ compactification manifold. Before discussing the
general $K_3$, we consider orbifolds and find that the amplitude
is  zero in such compactification limits. We also derive a general
expression valid for all $K_3$ compactifications. In Section 5, we
discuss the world-sheet degeneration limit relevant to large
radius compactifications and present explicit expressions for
theta functions, prime-form {\it etc}. As a first step toward
constructing more explicit and ``calculable'' examples, and in
order to gain more insight into the general case, we reconsider
orbifolds in Section 6, but now in the presence of non-vanishing
VEVs for the blowing-up modes. In Section 7, we overcome another
obstacle -- a cancellation due to the $\ZZ_2$ symmetry of the SS
mechanism, which occurs in the simplest form of SS
compactifications -- by considering its $\ZZ_N$ generalization \`a
la Kounnas and Porrati \kpor. This allows extracting the leading
radius dependence of the gaugino mass and thus its behavior for
small gravitino mass. Finally, Section 7 contains our conclusions
and outlook. Some technical details are relegated to three Appendices.

\newsec{Bordered $(g=1,h=1)$ Riemann Surface From the Involution of
$(g=2,h=0)$} In this section, we study the genus 3/2 Riemann
surface with one handle and one boundary, see Fig.1. The cycles
$\bf a$ and $\bf b$ shown in Fig.1 are the canonical homology
basis; the boundary $\bf c$ is homologous to $\bf ab^{-1}a^{-1}b$.
Such a surface can be obtained from the double torus of genus 2 by
applying the world-sheet involution $I$ that exchanges left and
right movers, according to Fig.2. Its action consists of a
reflection with respect to the plane of the dividing geodesics
$\bf c$, which in the canonical homology basis interchanges {\bf
a} and {\bf b} cycles, while simultaneously reverts the
orientation of the later~\invo: ${\bf a}_1\leftrightarrow{\bf
a}_2$ and ${\bf b}_1\leftrightarrow -{\bf b}_2$. It follows that
the involution symplectic matrix $I$, acting in order on blocks of
{\bf a} and {\bf b} cycles, takes the form: \eqn\involution{
I\equiv\left(\matrix{\Gamma&0\cr C& -\Gamma\cr}\right)=
\left(\matrix{\sigma^1&0\cr 0&-\sigma^1\cr}\right) \, ,} where
$\Gamma$ is a symmetric matrix of integers with $\Gamma^2=1$. In
our case, it is given by the $2\times 2$ Pauli matrix $\sigma^1$.
A period matrix $\Omega$ invariant under the involution
\involution\ satisfies~\invo:
${\bar\Omega}=I(\Omega)=(C-\Gamma\Omega)\Gamma=-\sigma^1\Omega\sigma^1$.
It can thus be put in the form \eqn\periodmatrix{
\Omega=\left(\matrix{\tau&-il\cr -il&-{\bar\tau}\cr}\right) \, ,}
where $\tau=\tau_1+i\tau_2$; it depends on three real parameters
$\tau_1$, $\tau_2$ and $l$. \fig{Bordered $g=1$ surface with $h=1$
boundary.}{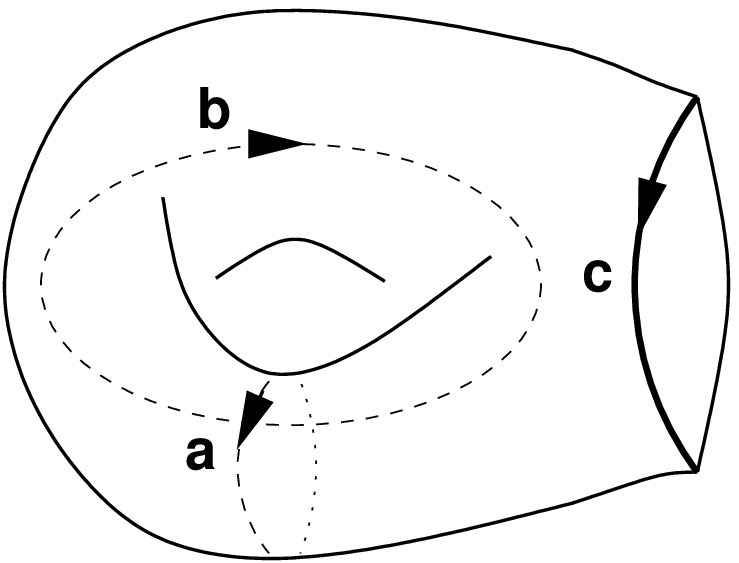}{3cm} \fig{The same $(g=1,h=1)$ surface
obtained by a mirror involution of $g=2$.}{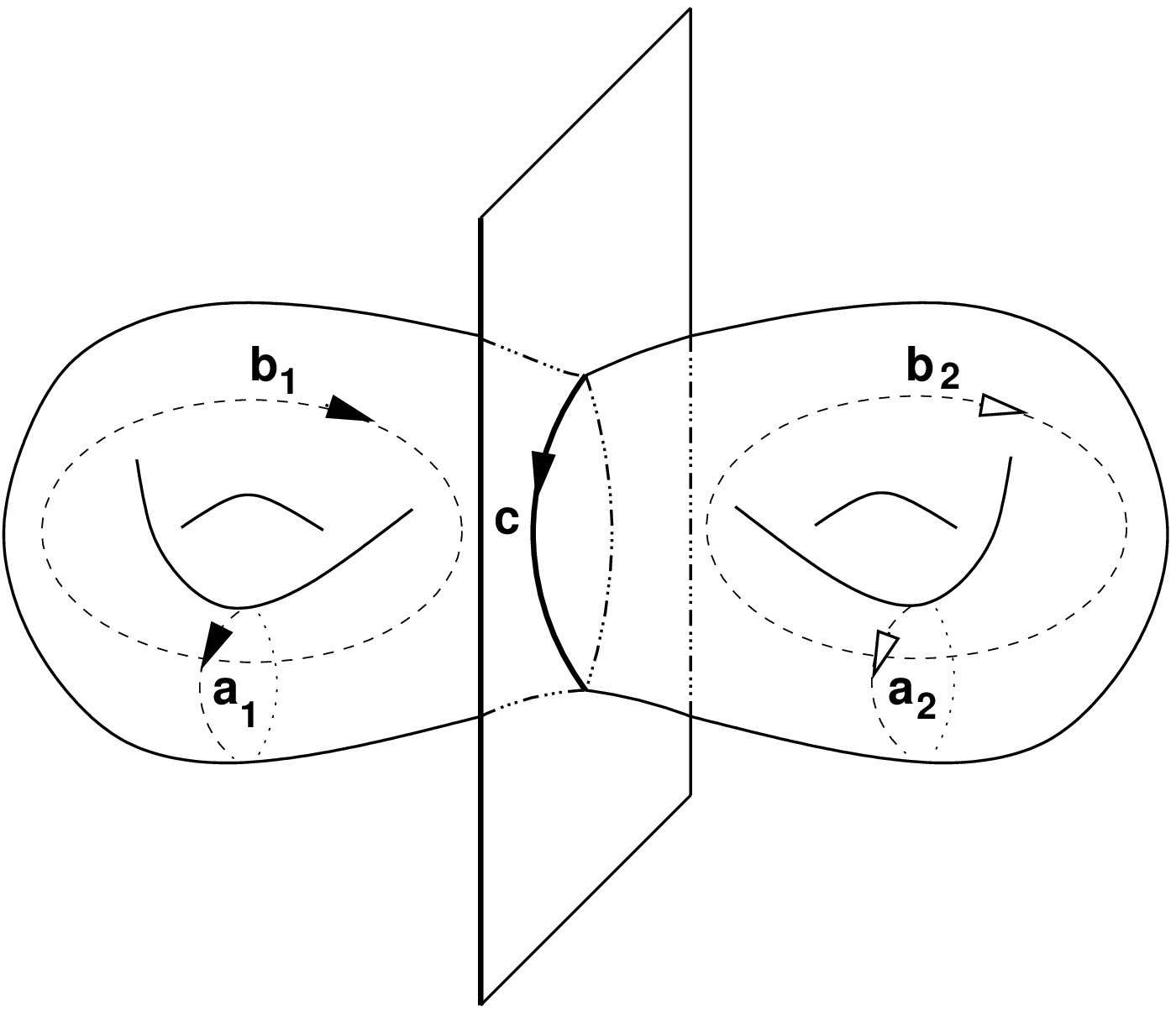}{6cm} The
``relative modular group," defined by the modular transformations
Sp(4, \bb{Z}) that preserve the involution, was studied in general
in Ref.~\modular. In the case under consideration, it consists of
the usual genus-one transformations that act simultaneously on the
two tori of the double cover. These are generated by the familiar
$S$ and $T$ transformations; $S$ interchanges {\bf a} with {\bf b}
cycles, up to a sign dictated by the involution: ${\bf
a}_1\leftrightarrow{\bf b}_1$ and ${\bf a}_2\leftrightarrow -{\bf
b}_2$; on the other hand, $T$ leaves invariant the {\bf a} cycles
and shifts the {\bf b} cycles by ${\bf b}_1\to{\bf b}_1+{\bf a}_1$
and ${\bf b}_2\to{\bf b}_2-{\bf a}_2$. In matrix notation, they
are given by \eqn\modtransf{ S=\left(\matrix{0&\sigma^3\cr
\sigma^3&0\cr}\right) \qquad ;\qquad T=\left(\matrix{1&0\cr
\sigma^3&1\cr}\right) \, ,} which, when applied to the period
matrix \periodmatrix, yield: \eqn\ST{ T:\ \tau\to\tau +1\ ,\ l\to
l\qquad ;\qquad S:\ \tau\to -{{\bar\tau}\over |\tau|^2-l^2}\ ,\
l\to {l\over |\tau|^2-l^2} \, ,} or equivalently: \eqn\STmatrix{
T:\ \Omega\to\Omega +\sigma^3\qquad ;\qquad S:\ \Omega\to
-\sigma^3\Omega^{-1}\sigma^3 \, .} Positivity of the period matrix
implies that $\tau_2$ and $-\det\Omega=|\tau|^2-l^2$ are positive.

The fundamental domain of integration can be easily derived from \ST. In
the $\tau$-plane, for fixed $l$, it consists of the part of the strip
$-1/2<\tau_1<1/2$ which lies above the circle of radius $\sqrt{1+l^2}$,
such that $|\tau|^2>1+l^2$ (see Fig.3).
Obviously, the parameter $l$ determines the size of the hole. In the
limit $l\to 0$, the surface becomes a regular torus and the modular
transformations \ST\ are reduced to the usual SL(2,\bb{Z})
transformations $\tau\to\tau +1$ and $\tau\to -1/\tau$ with their
familiar fundamental domain. Note that \ST\ preserve the sign of $l$
which remains undetermined. Actually, before the involution, there is no
constraint on the sign of the non-diagonal element of the period matrix
$\Omega_{12}$. However, after the involution, we show below that the sign
of $l$ is fixed to be positive~\ant.\fig{The shaded
region is the $(g=1,h=1)$ fundamental domain.}{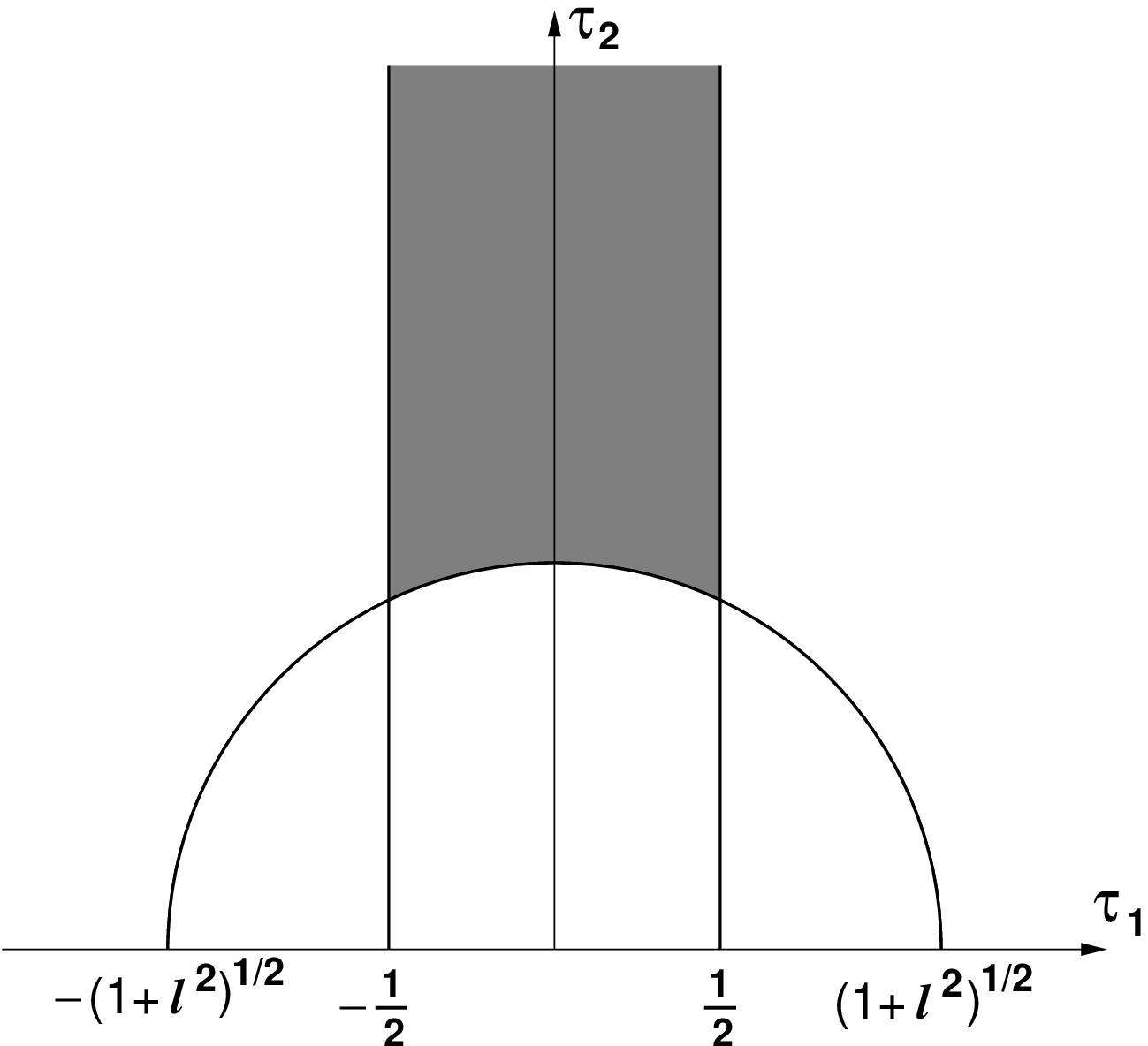}{6cm}

We use the basis of two holomorphic differentials $\omega_i$ on the
genus-two surface $\Sigma$, normalized to the {\bf a} cycles:
\eqn\omegas{
\int_{{\bf a}_j}\omega_i=\delta_{ij} \quad ;\quad
\int_{{\bf b}_j}\omega_i=\Omega_{ij} \quad ;\quad
\int_\Sigma\omega_i\wedge{\bar\omega}_j=\Im\Omega_{ij}
\, .}
Consider now the surface $\Sigma_1$ obtained from $\Sigma$ by the
involution \involution, and evaluate
$\int_{\Sigma_1}\omega_i\wedge{\bar\omega}_j$ using the identity~\Oog:
\eqn\ident{
\int_{\Sigma_1}\omega_i\wedge{\bar\omega}_j={i\over 2}\left(
\int_{{\bf a}_1}\omega_i\int_{{\bf b}_1}{\bar\omega}_j-
({\bf a}_1\leftrightarrow{\bf b}_1)\right)+
\int_{\bf c}\omega_i(z)\int^z{\bar\omega}_j
\, .}
{}From the involution \involution, one deduces that on the boundary
$\omega_i=\Gamma_{ij}{\bar\omega}_j$ and thus
\eqn\reality{{\bar\omega}_{1,2}(x)=\omega_{2,1}(x)\, ,\qquad x\in
{\bf c}\,.}
Using this property, one has
\eqn\property{
\int_{\Sigma_1}\omega_i\wedge{\bar\omega}_j=
\int_{\Sigma_1}\omega_i\wedge({\bar\omega}_j-\sigma^1_{jk}\omega_k)=
{i\over 2}\left( \int_{{\bf a}_1}\omega_i\int_{{\bf b}_1}
({\bar\omega}_j-\sigma^1_{jk}\omega_k)-
({\bf a}_1\leftrightarrow{\bf b}_1)\right)\, ,
}
since the boundary contribution vanishes. Using now the relations
\omegas\ and \periodmatrix, we find:
\eqn\integrals{
\int_{\Sigma_1}\omega_1\wedge{\bar\omega}_1=\tau_2-{l\over 2}\quad ;\quad
\int_{\Sigma_1}\omega_2\wedge{\bar\omega}_2={l\over 2}\quad ;\quad
\int_{\Sigma_1}\omega_1\wedge{\bar\omega}_2=-{l\over 2}
\, .}
{}From the second relation above, it then follows that $l$ is positive.

\newsec{Partition Functions}

In this section, we compute bosonic and fermionic partition functions, as
well as the SS deformation, on the genus 3/2 $(g=1,h=1)$ Riemann surface
that was described above. Following the analysis of Ref.~\invo, the
bosonic determinants can be obtained by taking the appropriate square root
of the corresponding expression on the genus 2 double cover, up to a
correction factor $R_\Sigma$ that depends on the involution and on the
boundary conditions, Neumann (N) or Dirichlet (D):\foot{Note the
difference in the definition of $R_\Sigma$ between the two references
of~\invo; the one is inverse of the other. Here, we use the definition of
Bianchi and Sagnotti.}
\eqn\bosdet{
g=2\ :\ (\det\Im\Omega)^{-1/2}|Z_1|^{-1}\quad \longrightarrow\quad
(g=1,h=1)\ :\ (R_{\Sigma_1}\det\Im\Omega)^{-1/4}Z_1^{-1/2}
\, ,}
where $Z_1$ is the chiral part of the determinant, and obviously after
the involution the period matrix $\Omega$ has the reduced form
\periodmatrix. The correction factor is given by
\eqn\corfac{
R_{\Sigma_1}^{\rm N}=(R_{\Sigma_1}^{\rm D})^{-1}=
\det\left({1-\Gamma\over 2}\Im\Omega+
{1+\Gamma\over 2}(\Im\Omega)^{-1}\right)=
{\tau_2 +l\over\tau_2 -l}
\, ,}
where $\Gamma$ is defined from the form of the involution \involution\
and equals $\sigma^1$ in our case. As a result, the bosonic determinant
on $\Sigma_1$ becomes $(\tau_2\pm l)^{-1/2}Z_1^{-1/2}$, where the plus
(minus) sign corresponds to Neumann (Dirichlet) boundary conditions. Note
that the correction factor \corfac\ is invariant under the modular
transformations \ST. It will be also useful to define
\eqn\tautilde{
{\tilde\tau}\equiv \tau_1 +i\sqrt{\tau_2^2-l^2}
\, ,}
which has the usual form of SL(2,\bb{Z}) transformations,
${\tilde\tau}\to{\tilde\tau}+1$ and ${\tilde\tau}\to -1/{\tilde\tau}$.

The generalization to a compact boson is straightforward. The
classical action  (see Appendix A) is: \eqn\Sclass{ S_{\rm
cl}={\pi R^2\over 2}({\vec m}+{\vec n}{\bar\Omega})
(\Im\Omega)^{-1}({\vec m}^T+\Omega{\vec n}^T) = {\pi R^2\over
(\tau_2\pm l)}\left( m^2 +2mn\tau_1 +n^2(|\tau|^2-l^2)\right) \,
,} where $R$ is the compactification radius in $\alpha^\prime$
units, and ${\vec m}=m(1,\mp 1)$ and ${\vec n}=n(1,\pm 1)$ are the
winding numbers around the ${\bf b}$ and ${\bf a}$ cycles,
respectively. Their form is fixed by the involution, while the
two signs correspond to N and D boundary conditions. Combining
\Sclass\ with the quantum determinant \bosdet, and summing over
all classical solutions, one finds the total contribution to the
partition function of a compact boson: \eqn\Zb{
Z_B=Z_1^{-1/2}\sum_{m,n}Z_{m,n}^{\rm N,D}\quad ;\quad Z_{m,n}^{\rm
N,D}={R\over (\tau_2\pm l)^{1/2}} e^{-\displaystyle\pi R^2\left(
{(m +n\tau_1)^2\over \tau_2\pm l} +n^2(\tau_2\mp l)\right)} \, .}
After performing a Poisson resummation in $m$, one can bring the
partition function $Z_{m,n}^{\rm N,D}$ into the Hamiltonian form
$Z_{{\tilde m},n}^{\rm N,D}$: \eqn\Poisson{ Z_{{\tilde m},n}^{\rm
N,D} =e^{\displaystyle{i\pi\over 2} \left(
p_L^2{\tilde\tau}-p_R^2{\bar{\tilde\tau}}\right) } \quad ;\quad
p_{L,R}={{\tilde m}\over {\tilde R}^{\rm N,D}} \pm n{\tilde
R}^{\rm N,D}\quad {\rm with}\quad {\tilde R}^{\rm N,D}\equiv
R/(R_{\Sigma_1}^{\rm N,D})^{1/4} \, ,} where ${\tilde\tau}$ was
defined in \tautilde. Thus, the lattice partition function takes
the familiar form in terms of an effective radius modified by the
correction factor \corfac. Obviously, in the limit where the size
of the hole vanishes, $l\to 0$, the correction factor becomes
unity and the partition function is reduced to its toroidal form.
{}From the expression \Poisson, it is also easy to see that
T-duality, $R\to 1/R$, exchanges N and D boundary conditions. Note
that $R\to 1/R$ is also formally equivalent to $l\to -l$.
\subsec{Fermions and Theta Functions}
  The method of taking the
square root from the double cover can be also applied to the
fermionic determinants giving rise to theta-functions. Each
complex fermion leads to 16 spin structures corresponding to four
boundary conditions ${\vec a}=(a_1,a_2)$ and ${\vec b}=(b_1,b_2)$
along the two non-trivial cycles ${\bf a}$ and ${\bf b}$ of
$\Sigma_1$ for the left and right movers: \eqn\deftheta{
\Theta\left[{{\vec a}\atop{\vec b}}\right](\Omega )= \sum_{{\vec
n}=(n_1,n_2)} e^{\displaystyle i\pi ({\vec n}+{\vec a})\Omega
({\vec n}+{\vec a})^T +2i\pi ({\vec n}+{\vec a})\sigma^3{\vec
b}^T} \, .} The insertion of $\sigma^3$ in the last phase was
chosen for convenience, so that in the toroidal limit $l\to 0$:
\eqn\factoriz{ l\to 0:\qquad \Theta\left[{{\vec a}\atop{\vec
b}}\right](\Omega )\longrightarrow \Theta\left({a_1\atop
b_1}\right)(\tau)\, {\bar\Theta}\left({a_2\atop
b_2}\right)({\bar\tau}) \, .} It follows that under the modular
transformations \STmatrix, $\Theta$ transforms as the product of
the 1-loop theta's $\Theta\bar\Theta$: \eqn\transftheta{\eqalign{
T:&\quad\Theta\left[{{\vec a}\atop{\vec b}}\right](\Omega
+\sigma^3)= e^{-i\pi a_1(a_1+1)+i\pi a_2(a_2+1)}\
\Theta\left[{{\vec a}\atop{\vec b}+{\vec a}+{1\over 2}}\right]
(\Omega )\, ,\cr S:&\quad\Theta\left[{{\vec a}\atop{\vec
b}}\right](-\sigma^3\Omega\sigma^3 ) =\sqrt{\det\Omega}\
e^{2i\pi(a_1b_1-a_2b_2)} \Theta\left[{-{\vec b}\atop{\vec
a}}\right](\Omega ) \, .\cr}} The partition function involves a
sum over spin structures with appropriate coefficients
$c\left[{{\vec a}\atop{\vec b}}\right]$, determined at one loop
level. As usually, at higher loops the corresponding coefficients
are determined by the factorization properties of the vacuum
amplitude. In our case, it is sufficient to consider the $l\to 0$
limit, to deduce in a shorthand notation: \eqn\coefs{
c\left[{{\vec a}\atop{\vec b}}\right]=c_L\left({a_1\atop
b_1}\right)\ c_R\left({a_2\atop b_2}\right) \, ,} where the
subscripts $L,R$ denote the left and right movers in the torus
amplitude.

\subsec{The Scherk-Schwarz Deformation and the Large Radius Limit}

Here, we will consider the breaking of bulk supersymmetry via SS
boundary conditions along a direction transverse to the D-brane
stack \refs{\SS,\rt,\kpor,\SSII,\SSI}. In the closed string
sector, it amounts to deform the partition function by coupling
the lattice momenta to the fermionic spin structures, imposing
antiperiodic boundary conditions to space-time fermions
consistently with world-sheet modular invariance: \eqn\ssdef{
\Theta\left[{{\vec a}\atop{\vec b}}\right]\longrightarrow
e^{-2i\pi n(b_1-b_2)}\ \Theta\left[{{\vec a}+n{\vec 1}\atop{\vec
b}+m{\vec 1}}\right]= \delta_s^{\rm SS}\ \Theta\left[{{\vec
a}\atop{\vec b}}\right]\quad ;\quad \delta_s^{\rm
SS}=e^{2i\pi\left[ m(a_1-a_2)-n(b_1-b_2)\right] } \, ,} where
${\vec 1}\equiv (1,1)$, and $m,n$ are the winding numbers
appearing in $Z_{m,n}^D$ of \Zb. The choice of Dirichlet
conditions follows from the transversality of the SS direction to
the boundary. Thus, the SS deformation consists of inserting the
phase $\delta_s^{\rm SS}$ \ssdef\ in the partition function. In
the zero winding sector, $n=0$, upon Poisson resummation in $m$,
it amounts to shift the Kaluza-Klein momenta ${\tilde m}$ in
\Poisson\ of space-time fermions $(a_1-a_2=1/2)$ by 1/2, giving a
mass to the gravitino $1/2R$. In the odd winding $n$ sector, on
the other hand, there is a change in the GSO projection due to the
additional phase.

We consider now the large radius limit $R\to\infty$, where supersymmetry
is restored. The relevant partition function is $Z_{m,n}^D$,
corresponding to the lower sign of \Zb. Since at the origin of the
lattice, $m=n=0$, the SS phase becomes trivial, the contribution to the
gaugino mass should vanish by supersymmetry. Thus, for finite moduli
($\tau_2$ and $l$) the result is exponentially suppressed in $R$. To
avoid exponential suppression, $n$ should vanish and $\tau_2$ should
go to infinity. This limit corresponds to the handle degeneration,
describing a massless closed string exchange in the loop, and thus
gravitational corrections in the effective field theory.

Note that for Neumann boundary conditions, one should get an additional
contribution from all winding numbers $n\ne 0$ in the limit
$\tau_2, l\to\infty$ with $\tau_2 -l\to 0$, so that $\tau_2^2-l^2$
remains fixed. This degeneration describes precisely the
non-gravitational two-loop non-planar diagram in the gauge theory of the
D-branes. Obviously, this limit is not relevant for our purposes, since
it does not contain any information on the mediation of supersymmetry
breaking from the bulk to the boundary.

\newsec{Gaugino Mass as a Descendant of a Topological Amplitude}

In this section, we evaluate the two-point function involving two
gauginos at zero momentum coupled to the boundary of a $(g=1,h=1)$
surface $\Sigma_1$. We consider the case with $N=2,~ D=4$
supersymmetry on the boundary. Thus the $D=10$ type II theory is
compactified on the six-dimensional internal space $K3\times T^2$,
where $T^2$ contains the Scherk-Schwarz circle transverse to the
boundary direction. The $K3$ manifold is described by a (4,4)
superconformal field theory, yielding $N=4$ supersymmetry in the
bulk, spontaneously broken by the SS boundary conditions. As it
becomes clear in the following, the computation of the gaugino
mass is closely related to the computation of the genus 2
topological amplitude ${\cal F}_2$ that determines the
gravitational coupling $W^4$ in Calabi-Yau compactifications of
type II theory \agnt. Before a full-fledged discussion of $K3$, it
is very instructive to consider its orbifold limit, with $K3$
realized as $T^4/H$, where $H=\ZZ_2$ reflection $(z\to -z)$ or
another $\ZZ_N$ symmetry of $T^4$.

\subsec{The Orbifold Case}

The gaugino vertex operator of definite chirality $\alpha$, at zero
momentum, in the canonical $-1/2$ ghost picture, reads:
\eqn\vertex{
V^{(-1/2)}_\alpha(x)=:e^{-\varphi/2}S_\alpha S_{int}:
\, ,}
where $x$ is a position on the boundary of the world-sheet, $\varphi$ is
the scalar bosonizing the superghost system, and $S_\alpha$ ($S_{int}$) is
the space-time (internal) spin field. Upon complexification of the four
space-time fermionic coordinates, $\psi_I$ for $I=1,2$, and introducing the
bosonized scalars $\psi_I=e^{i\phi_I}$, one has
\eqn\spinf{
S_\alpha =e^{\pm\displaystyle {i\over 2}(\phi_1 +\phi_2)}
\quad ;\quad\alpha=\pm
\, .}
Similarly, bosonizing the fermionic coordinates of $T^2$ and $K3$, by
introducing the scalars $\phi_3$ and $\phi_{4,5}$, respectively, one has:
\eqn\spinfint{
S_{int} =e^{\displaystyle {i\over 2}(\phi_3 +\phi_4 +\phi_5)}
\, .}

In order to compute the amplitude involving two fermions \vertex\
at the boundary of a \break $(g=1, h=1)$ surface, one has to
insert three picture changing operators $e^\varphi T_F$, where
$T_F$ is the world-sheet supercurrent. We choose to insert all
these operators on the boundary $\bf c$. One of them is needed to
change the ghost-picture of one gaugino to $+1/2$, while the other
two arise from the integration over the supermoduli: \eqn\amp{
m_{1/2}=g_s^2\int_{F(\Sigma_1)}
d\mu(\Omega)\int_{\partial\Sigma_1}
      dxdy\ {\cal A}\quad ;\quad {\cal A}=
\left\langle V^{(-1/2)}_+(x)V^{(-1/2)}_-(y)\prod_{a=1}^3
e^\varphi T_F(z_a)\right\rangle
\, ,}
where $g_s$ is the string coupling. Its square includes $g_s$ associated
to Euler characteristic $-1$ as well as an extra $g_s$ from
the normalization of gaugino kinetic terms on the disk. The moduli
integration is over the fundamental domain $F(\Sigma_1)$, shown in Fig.~3,
with the appropriate measure $d\mu(\Omega)$. The dependence on
the positions $z_a$ of the picture changing operators is gauge artifact
and should disappear from the physical amplitude.

By internal charge
conservation, since both gauginos carry charge $+1/2$ for $\phi_3$,
$\phi_4$ and $\phi_5$, only the internal part of the world-sheet
supercurrents, $T_F^{int}$ contributes; each $T_F$ should provide $-1$
charge for $\phi_3$, $\phi_4$ and $\phi_5$, respectively. In the orbifold
case, $T_F^{int}=\sum_{I=3}^5\psi_I^*\partial X^I+ c.c.$, where $X^3$ and
$X^{4,5}$ are the complexified coordinates of $T^2$ and $K3$,
respectively. The later transform under an element $h$ of the orbifold
group $H$ as $X^4\to hX^4$ and $X^5\to h^{-1}X^5$.
The amplitude \amp\ then becomes:
\eqn\ampA{
{\cal A}=\langle e^{-\varphi/2}(x)e^{-\varphi/2}(y)\!
\prod_{I=3}^5\! e^\varphi(z_I)\rangle\!
\prod_{I=1}^2\! \langle e^{i\phi_I/2}(x)e^{-i\phi_I/2}(y)\rangle\!
\prod_{I=3}^5\! \langle e^{i\phi_I/2}(x)e^{i\phi_I/2}(y)e^{-i\phi_I}(z_I)
\rangle ,}
where $\{ z_I\}$, $I=3,4,5$, is a permutation  of $\{ z_a\}$, $a=1,2,3$, and
an implicit summation over all permutations is understood.

Performing the contractions for a given spin structure $s$, one finds:
\eqn\ampAs{\eqalign{
{\cal A}_s= ~&
{\theta_s^2({1\over 2}(x-y))\prod_{I=3}^5\theta_{s,h_I}
({1\over 2}(x+y)-z_I)\,\partial X_{h_I}(z_I)\over
\theta_s({1\over 2}(x+y)-\sum_{I=3}^5z_I+2\Delta)}\cr
~&\times{\sigma(x)\sigma(y)\over\prod_{I<J}^{3,4,5}E(z_I,z_J)
\prod_{I=3}^5\sigma^2(z_I)}\times{Z_2\over Z_1^4
\prod_{I=3}^5Z_{1,h_I}}\, \delta_s^{\rm SS}\, Z_{lat}\, ,}}
where $\theta_s$ is the genus-two theta-function of spin structure $s$,
$E$ is the prime form, $\sigma$ is a one-differential with no zeros or
poles and $\Delta$ is the Riemann $\theta$-constant \refs{\ver,\fay}.
$Z_{1,h_I}$ is the (chiral) non-zero mode determinants of the $h_I$-twisted
$(1,0)$ system, with $h_3=1,\, h_4=h,\, h_5=h^{-1}$, while
$Z_1\equiv Z_{1,1}$, and
$Z_2$  is the chiral non-zero mode determinant of the $(2,-1)$
{\it b-c} ghost system. Finally, $Z_{lat}$ stands for all zero-mode
parts of space-time and internal coordinates. In our case, it is given by:
\eqn\zrest{
Z_{lat}={1\over (\tau_2 +l)^2}\,
Z_{m,n}^D\, Z_{rest}
\, ,}
where an implicit summation over $m,n$ should be performed, taking into
account also $\delta_s^{\rm SS}$ and the $\partial X_I$ factors in
\ampAs. Recall that $\delta_s^{\rm SS}$ is the phase \ssdef\ of the SS
deformation, $Z_{m,n}^D$ is the momentum lattice of the SS circle \Zb,
satisfying Dirichlet conditions at the boundary, while $Z_{rest}$ denotes
the remaining zero-mode contribution of the $K3$ lattice,
together with the additional (non-SS) direction of $T^2$.

The large radius limit along the SS direction was described in the
previous section. In this limit, the amplitude is exponentially
suppressed unless  $\tau_2\to\infty$ with $l$ fixed, corresponding to
the effective field theory gravitational loop exchange. Moreover, the
momentum sum is restricted to vanishing winding number $n$. Then as
\eqn\zerstlim{
R\to\infty:\quad\delta_s^{\rm SS} =  (-1)^{2m\vec{a}\cdot\vec{1}}\qquad
Z_{lat}\sim
{1\over(\tau_2 +l)^2}\,
Z_{m, 0}^D\, Z_{rest}(\tau_2\to\infty)
\, .}
Note that for $m$ even, the SS deformation becomes trivial hence, as
shown below,
the result vanishes by supersymmetry; the SS lattice sum
is in fact restricted to the odd winding numbers $m$.

In order to perform an explicit sum over spin structures,
we choose the positions of the picture-changing operators
satisfying the condition \agnt
\eqn\gauged{\sum_{I=3}^{I=5}z_I=y+2\Delta\, .}
Then the
spin structure-dependent part of the amplitudes \ampAs\ simplifies, yielding
the following sum: \eqn\ssorbi{{\cal S_O}=
\sum_{\vec a,\vec b}(-1)^{2m\vec{a}\cdot\vec{1}}
\theta\left[{{\vec a}\atop{\vec b}}\right]
({1\over 2}(x-y))\prod_{I=3}^5
\theta_{h_I}\left[{{\vec a}\atop{\vec b}}\right]
({1\over 2}(x+y)-z_I)\, ,}
where the theta-function subscripts $h_I$ denote additional shifts in their
characteristics associated to to the respective orbifold group elements.
By using the Riemann identity \fay, we obtain
\eqn\ssorbj{{\cal S_O}=
\theta\left[{0\atop {m\over 2}\vec 1}\right](x-\Delta)
\prod_{I=3}^5\theta_{h_I^{-1}}\left[{0\atop {m\over 2}\vec 1}\right]
(z_I-\Delta)\, .}
Note that as anticipated, for $m$ even, the above result vanishes
by supersymmetry,
due to the Riemann
vanishing theorem, $\theta(z-\Delta)=0 ~\forall
z\in\Sigma$. For $m$ odd, the amplitude becomes
\eqn\ampAso{\eqalign{
{\cal A}= ~&\theta\left[{0\atop {1\over 2}\vec 1}\right](x-\Delta)
{\sigma(x)\sigma(y)\over\prod_{I<J}^{3,4,5}E(z_I,z_J)
\prod_{I=3}^5\sigma^2(z_I)}\times{Z_2\over Z_1^4\prod_{I=3}^5Z_{1,h_I}}\cr
~& \times\prod_{I=3}^5
\theta_{h_I^{-1}}\left[{0\atop {1\over 2}\vec 1}
\right](z_I-\Delta)\,\partial X_{h_I}(z_I)\times Z_{lat}\, .}}

As mentioned before, the amplitude \ampAso\ contains an implicit
sum over 6 permutations $\{z_I\}$ of the supercurrent insertion
points $\{z_a\}$. Due to complete antisymmetry of the prefactor,
this leads to the antisymmetrization of the product
$\prod_{a=1}^3\theta_{h_I^{-1}}(z_a-\Delta)\partial X_{h_I}(z_a)$
in $\{z_a\}$. Recall that $h_3=1,\, h_4=h,\, h_5=h^{-1}$. {}For a
$T^4/\ZZ_2$ orbifold, $h_4=h_5$, hence the amplitude vanishes
after summing over all permutations. In fact, as shown in Section
6, it also vanishes for all $T^4/\ZZ_N$ orbifolds. This
cancellation is clearly due to discrete symmetries, therefore it
is not expected to hold for a generic $K_3$ compactification.

\subsec{The General $K3$ Case}

The case of a general $K3\times T^2$ compactification can be discussed by
slightly modifying most of previous computations. The underlying $N=4$
superconformal field theory (SCFT) of central charge ${\hat c}=4$ is
characterized by an $SU(2)$ affine algebra at level one. This can be
realized in terms of a free boson $\Phi$ compactified at the self-dual
radius, that couples also to the spin structure $s$. The internal part
\spinfint\ of the gaugino vertex operator \vertex\ becomes:
\eqn\spinfintK{
S_{int}=e^{i\phi_3/2+i\Phi/{\sqrt 2}}
\, ,}
while the internal part of the world-sheet supercurrent reads:
\eqn\TFint{
T_F^{int}=\psi_3^*\partial X^3+e^{-i\Phi/{\sqrt 2}}{\hat G}_- +c.c.
\, ,}
where ${\hat G}_-$ has no singular operator product expansion with the
$U(1)$ current $\partial \Phi$ and carries a conformal dimension 5/4.
The internal part of the amplitude corresponding to the last product of
contractions in \ampA\ then becomes:
\eqn\ampAint{\eqalign{
{\cal A}_{int}= ~& \langle
e^{i\phi_3/2}(x)e^{i\phi_3/2}(y)e^{-i\phi_3}(z_3)\rangle
\langle e^{i\Phi/{\sqrt 2}}(x)e^{i\Phi/{\sqrt 2}}(y)
e^{-i\Phi/{\sqrt 2}}(z_4)e^{-i\Phi/{\sqrt 2}}(z_5)\rangle\cr
&\hskip 1cm\times\langle{\hat G}_-(z_4){\hat G}_-(z_5)\rangle
\, ,\cr}}
giving rise to the following spin structure dependent amplitude:
\eqn\ampAsK{\eqalign{
{\cal A}_s= ~&
{\theta_s^2({1\over 2}(x-y))\theta_{s}({1\over 2}(x+y)-z_3)
Ch_{r,s}(x+y-z_4-z_5)
\over
\theta_s({1\over 2}(x+y)-\sum_{I=3}^5z_I+2\Delta)}
{E^{1/2}(z_4,z_5)\sigma(x)\sigma(y)\over\prod_{I<J}^{3,4,5}E(z_I,z_J)
\prod_{I=3}^5\sigma^2(z_I)}\cr
~&\times{Z_2\over Z_1^4}\, \delta_s^{\rm SS}\,
\partial X_3(z_3)\, Z_{lat\, T^2}
\langle{\hat G}_-(z_4){\hat G}_-(z_5)\rangle_r
\, .\cr}}
Here, the label $r=0,1/2$ denotes the two different types of spectral
flow of the $N=4$ SCFT, and a summation over $r$ is implicit. $Z_{lat\,
T^2}$ contains the zero-mode parts in the partition function of
space-time and $T^2$ coordinates, while the $K3$ contribution is included
in the last factor which is also independent of the spin structure.
Finally, $Ch_{r,s}(\nu)$ is the spin structure dependent $SU(2)$ level
one character:
\eqn\chrs{\eqalign{
Ch_{\vec r}\left[{{\vec a}\atop{\vec b}}\right]({\vec v})=~& Z_1^{-1/2}
\sum_{\vec n} e^{\displaystyle 2i\pi
({\vec n}+{\vec r}+{\vec a})\Omega ({\vec n}+{\vec r}+{\vec a})^T
+ 2i\pi ({\vec n}+{\vec r}+{\vec a})
\sigma^3({\vec v}+2{\vec b})^T}\cr ~&
=Z_1^{-1/2}(-1)^{4\vec b\cdot (\vec r+\vec a)}\theta
\left[{{\vec r+\vec a}\atop 0}\right](2\Omega,\vec v)\, .}}

In order to perform an explicit sum over spin structures,
we choose here the positions of the picture-changing operators
satisfying the same condition as in \gauged.
Then the spin structure-dependent part of the amplitude \ampAsK\
simplifies, yielding the following sum:
\eqn\sssum{\eqalign{{\cal
S_{K}}_3=~&\sum_{\vec a,\vec b} (-1)^{2m\vec a\cdot\vec 1}(-1)^{4\vec
b\cdot (\vec r+\vec a)}\cr&\times
\theta\left[{{\vec a}\atop{\vec b}}\right]({1\over 2}(x-y))\theta
\left[{{\vec a}\atop{\vec b}}\right]({1\over 2}(x+y)-z_3)
\theta \left[{{\vec r+\vec a}\atop 0}\right](2\Omega,x+y-z_4-z_5)\, ,}}
where we used the SS factor
$\delta_s^{\rm SS}=(-1)^{2m\vec a\cdot\vec 1}$ relevant to
the $R\to\infty$ limit, Eq.\zerstlim.
Now the summation over $\vec b$ can be performed by using the ``addition''
theorem \fay:
\eqn\addt{\eqalign{
{1\over 2^g}\sum_{\epsilon}(-1)^{4\vec\alpha\cdot\vec\epsilon}~&\theta
\left[{{\vec \alpha
+\vec\beta}\atop{\vec \gamma+
\vec\epsilon}}\right](u_1+u_2)\;\theta\left[{{\vec \alpha
+\vec\beta}\atop{\vec\epsilon}}\right](u_1-u_2)\cr &\hskip 1cm =
\theta\left[{{\vec \alpha}\atop{\vec\gamma}}\right](2\Omega,2u_1)
\,\theta\left[{{\vec \beta
+\vec\gamma}\atop{\vec\epsilon}}\right](2\Omega,2u_2)\, ,}}
with the result
\eqn\sssun{{\cal S_{K}}_3=\sum_{\vec a}\delta_s^{\rm SS}
\theta \left[{{\vec r+\vec a}\atop 0}\right](2\Omega,x+y-z_4-z_5)\,
\theta \left[{{\vec r+\vec a}\atop 0}\right](2\Omega,x-z_3)\,
\theta \left[{{\vec r}\atop 0}\right](2\Omega,y-z_3)\, .}
Finally, after using the inversion of \addt\ \fay,
we obtain
\eqn\sumfin{ {\cal S_K}_3=(-1)^{2m\vec r\cdot\vec 1}\,
\theta \left[{0\atop {m\over 2}\vec 1}\right](x-\Delta)\,
\theta \left[{0\atop {m\over 2}\vec 1}\right](z_3-\Delta)\,
\theta \left[{{\vec r}\atop 0}\right](2\Omega,z_4+z_5-2\Delta)}
As expected, the above result vanishes by supersymmetry
for $m$ even, due to the Riemann
vanishing theorem, $\theta(z-\Delta)=0 ~\forall
z\in\Sigma$. For $m$ odd, the amplitude becomes
\eqn\ampAL{\eqalign{
{\cal A}= ~&(-1)^{2\vec r\cdot\vec 1}\theta
\left[{{\vec r}\atop 0}\right](2\Omega,z_4+z_5-2\Delta)
\langle{\hat G}_-(z_4){\hat G}_-(z_5)\rangle_r
E^{1/2}(z_4,z_5)\sigma(z_4)\sigma(z_5)\, \cr ~&\times
\partial X_3(z_3)
\widetilde\omega(z_3)
{\widetilde\omega(x)\sigma(y)\over\prod_{I<J}^{3,4,5}E(z_I,z_J)
\prod_{I=3}^5\sigma^3(z_I)}\times{Z_2\over Z_1^{9/2}}Z_{lat\, T^2}\, ,}}
where
\eqn\omtil{\widetilde\omega(x)\equiv{1\over 2\pi i}\theta
\left[{0\atop {1\over 2}\vec 1}\right](x-\Delta)\sigma(x)}
is the holomorphic one-differential twisted by $(-1)$ along the
two $\bf b$-cycles
of the genus 2 double-cover,
{\it i.e.}\ twisted by $(-1)$ along the $\bf b$-cycle of the bordered surface.
In Eq.\ampAL, and in most cases below,
we neglect some purely numerical ({\it i.e.}\/ position-
and moduli-independent) factors.

At this point, the amplitude \ampAL\ still seems to depend
of the positions $\{z_a\}$ of the supercurrent insertions. However,
it contains an implicit sum over 6 permutations $\{z_I\}$ of these points.
As a result, the factor
\eqn\bfac{\eqalign{
B(z_I)\equiv ~&(-1)^{2\vec r\cdot\vec 1}\theta
\left[{{\vec r}\atop 0}\right](2\Omega,z_4+z_5-2\Delta)
\langle{\hat G}_-(z_4){\hat G}_-(z_5)\rangle_r
E^{1/2}(z_4,z_5)\sigma(z_4)\sigma(z_5) \cr ~&\hskip 2cm\times
\partial X_3(z_3)
\widetilde\omega(z_3)}}
yields a completely antisymmetric
combination $B(z_a)$ that transforms as a quadratic differential in each $z_a$,
twisted by $(-1)$ along the two
$\bf b$-cycles, with first order zeroes  as $z_a\to z_b$.
This implies that
\eqn\bfad{B(z_a)=B\det\widetilde{h}_a(z_b)\, ,}
where $B$ is constant (position independent) and $\widetilde{h}_a$ are the
zero modes  of the twisted $(2,-1)$ {\it b-c}
system. For the latter, we can use the bosonisation formula \refs{\ver,\russo}
\eqn\bosver{\theta
\left[{0\atop {1\over 2}\vec 1}\right](\sum_{a=1}^{a=3} z_a-3\Delta)
\prod_{a<b}^{1,2,3}E(z_a,z_b)
\prod_{a=1}^{a=3}\sigma^3(z_a)Z_1^{-1/2}=
\det\widetilde{h}_a(z_b)\widetilde{Z}_2\, ,}
where $\widetilde{Z}_2$ is the nonzero-mode determinant of the twisted
{\it b-c} system. Finally, after using all these relations, together
with \gauged,
we obtain:
\eqn\ampAM{
{\cal A}= {Z_2\over\widetilde{Z}_2 Z_1^5}BZ_{lat\, T^2}\;
\widetilde\omega(x)\,\widetilde\omega(y)\, ,}
which does {\it not\/} depend on the supercurrent insertion points.

In order to compute the gaugino mass, the amplitude \ampAM\ should
be integrated over the positions of the gaugino vertex operators
and over the moduli of $\Sigma_1$. Since $x,y\in {\bf c}\sim {\bf
ab^{-1}a^{-1}b}$, \eqn\omint{\int_{\bf c}\widetilde\omega(x)=2
\int_{\bf a}\widetilde\omega(x)\, .} We show in the Appendix that,
with the twisted differential normalized as in \omtil,
\eqn\limo{\lim_{\tau_2\to \infty}\int_{\bf
a}\widetilde\omega(x)=1\, .} In this way, we obtain
\eqn\mgu{m_{1/2}= g_s^2\int d\mu(\Omega){Z_2\over\widetilde{Z}_2
Z_1^3} BZ_{lat\, T^2}\, ,} which still remains to be integrated
over the moduli of $\Sigma_1$. In the next Section and in
Appendices A, B and C, we discuss the degeneration limit
$\tau_2\to\infty$, and derive explicit expressions for the
integration measure and the determinants.

\newsec{Degeneration Limit $\tau_2\to\infty$, $l$ {\rm fixed}}

In the limit $\tau_2\to\infty$, $l$ fixed, the two $\bf a$-cycles
of the double cover are ``pinched'' along the handles, see Fig.4.
The limiting curve $\Sigma_{\tau}$ is a sphere with two pairs of
points, $(a,b)$ and $(\bar a,\bar b)$, identified as the
punctures. This curve can be constructed from a two-sphere by
cutting out small disks on its surface (with diameters of order
$e^{-\pi\tau_2}$), and applying the ``plumbing'' procedure for
attaching the handles \fay. On the projective plane \P$^1$, the
handles extend symmetrically from $a$ to $b$ and from $\bar a$ to
$\bar b$, see Fig.5. The boundary of $\Sigma_{1\tau}$ is mapped
to the real axis, $\Im\, z=0$. The $\bf a$-cycle winds around
point $a$ while the $\bf b$-cycle runs from $b$ to $a$.
\fig{The $\tau_2\to\infty$, $l$ fixed degeneration limit of the double-cover}
{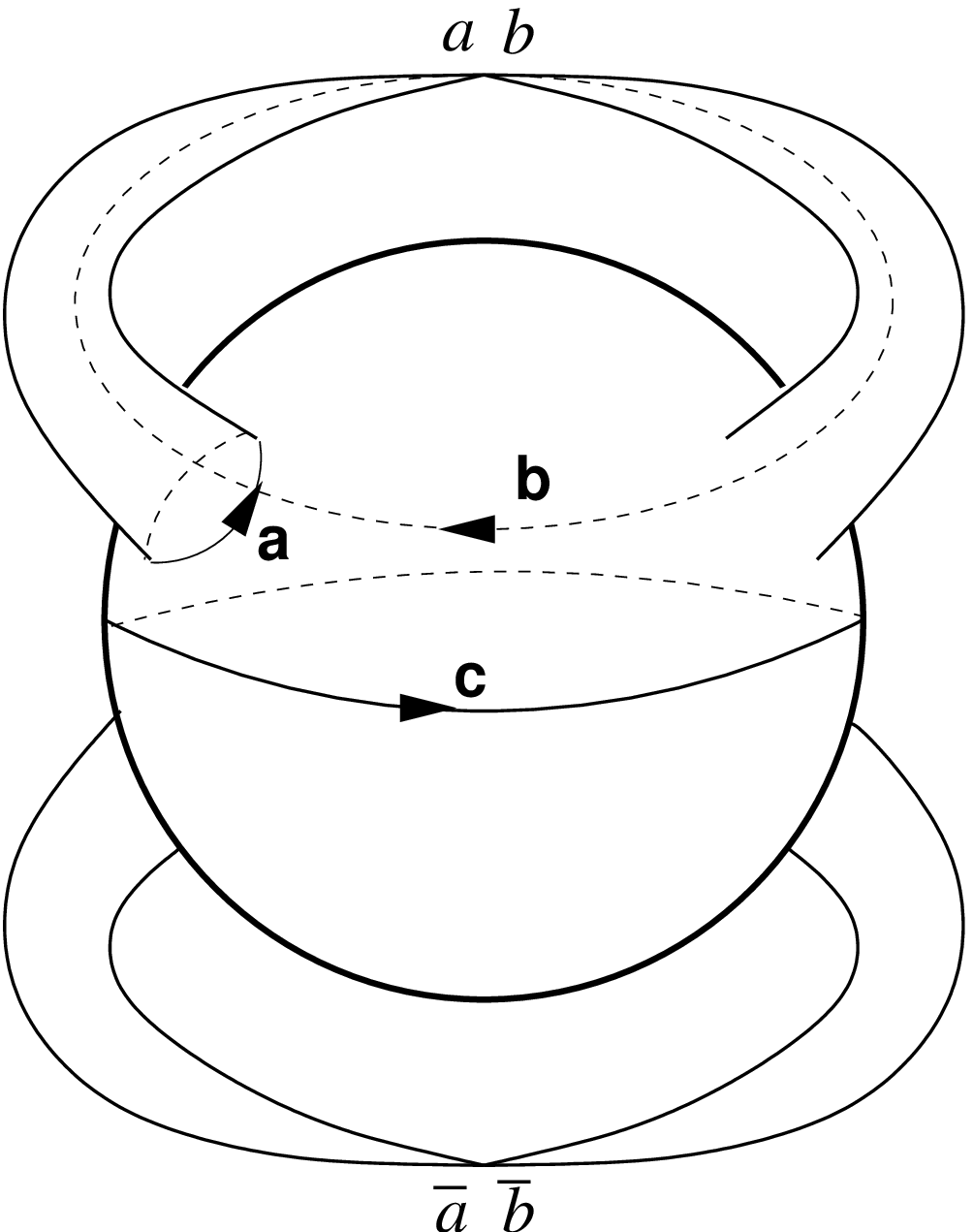}{4cm}
\fig{Degenerate surface on the projective plane.}{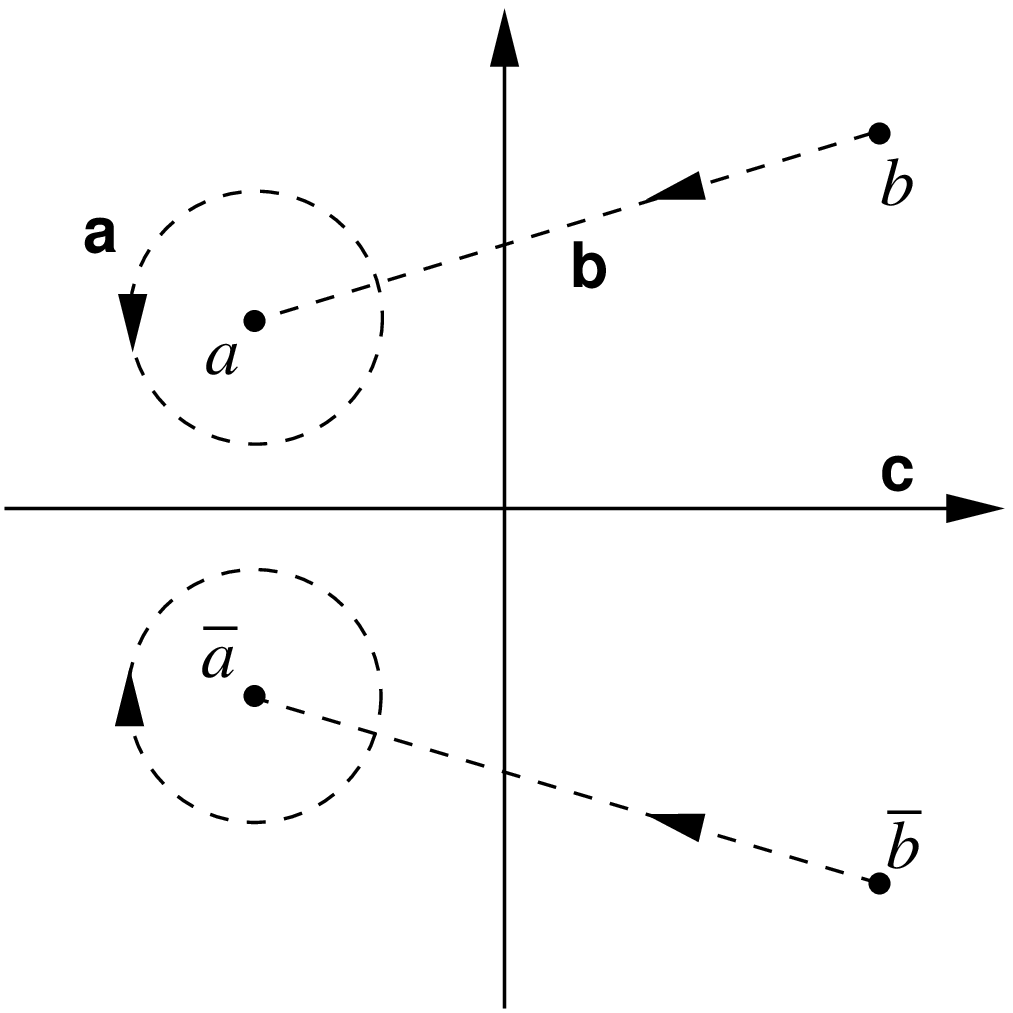}{4cm}

On $\Sigma_{\tau}$, the normalized basis of holomorphic one-forms
is given by \fay:
\eqn\hdif{
\omega_1= (z\, ;a,b)~, \qquad\qquad
\omega_2= (z\, ;\bar b,\bar a)~ ,} where the $SL(2,C)$-covariant
one-form \eqn\formdef{(z\, ; p_1,p_2)\equiv {1\over 2\pi
i}\left({1\over z-p_1}-{1\over z-p_2}\right)\, .
}
Note that on the
real axis, $\bar \omega_{1,2}=\omega_{2,1}$, as required by the
involution \involution. Furthermore, Eq.\hdif\ holds up to the
order $O(e^{-\pi\tau_2})$. Thus in the leading $O(1)$ order, the
generalized Jacobi map becomes \eqn\jac{z\in{\rm \P}^1
~\longrightarrow~ e^{2\pi iz_1}={z-a\over z-b}\cdot c_1(p)~,~
e^{2\pi iz_2}={z-\bar b\over z-\bar a}\cdot c_2(p)~,} where the
multiplicative constants $c_1$ and $c_2$ depend on the base point
$p$. They will cancel inside all theta function arguments,
therefore from now on we can set $c_1(p)=c_2(p)=1$. The
off-diagonal period matrix element is determined by the
$SL(2,C)$-invariant cross-ratio of the four puncture points:
\eqn\loff{e^{2\pi l}=(a,b\, ;\bar a,\bar b)= { (a-\bar b)(b-\bar a)
\over(a-\bar a)(b-\bar b)}.} Note that as expected, $l>0$,
since $a$ and $b$ lie on the same half-plane of \P$^1$.

Naively, the $\tau_2\to\infty$ limit of the theta function seems
to be trivial [$\theta(\Omega,z)\to 1$], unless one considers an
argument that diverges in this limit. Indeed, the arguments of the
form $z-\Delta$ that we are interested in, have this property
because $\Delta_i\propto-\Omega_{ii}/2\to\infty$. It is convenient
to write \eqn\zd{z-\Delta=z-\Delta_{\delta}+\delta(\Omega),} where
the half-period \eqn\halfp{\delta(\Omega)=\half\Omega\vec{1}
=\half\left({\tau-il\atop-\bar\tau-il}\right)} and
\eqn\dd{\Delta_{\delta}=\Delta+\delta(\Omega)} is the {\it finite}
Riemann constant associated to the spin structure $\delta$. In the
leading order of the theta series expansion,
\eqn\thss{\theta(z+\delta)=1+e^{-\pi l}e^{-2\pi iz_1} +e^{-\pi
l}e^{-2\pi iz_2}+e^{-2\pi i(z_1+z_2)}\,  ,} with a remainder of
order $O(e^{-\pi\tau_2})$. With the Jacobian map given in \jac,
we obtain \eqn\ths{\theta(z-\Delta)=1+e^{-\pi l}e^{2\pi
i\Delta_{\delta_1}} {z-b\over z-a}+e^{-\pi l}e^{2\pi
i\Delta_{\delta_2}}{z-\bar a\over z-\bar b} +e^{2\pi
i(\Delta_{\delta_1}+\Delta_{\delta_2})}{(z-b)(z-\bar a) \over
(z-a)(z-\bar b)}\, .} Then the Riemann vanishing theorem,
$\theta(z-\Delta)=0$, $\forall z\in {\rm\P}^1$, dictates
\eqn\divisors{e^{2\pi i\Delta_{\delta_1}}=-e^{-\pi l}\,{a-\bar
b\over b-\bar b}~, \qquad e^{2\pi i\Delta_{\delta_2}}=-e^{-\pi
l}\,{a-\bar b\over a-\bar a}~,} with $l$ determined by Eq.\loff.

{}For the prime-form, the limit can be taken in the same way as for the
theta function, and yields the obvious result:
\eqn\prif{E(x,y)=x-y\, .} The differential $\sigma(z)$ is more
subtle. One can start from the quotient \ver
\eqn\sz{{\sigma(z)\over\sigma(w)}={\theta(z-p_1-p_2+\Delta)\over
\theta(w-p_1-p_2+\Delta)} {E(w,p_1) E(w,p_2)\over
E(z,p_1)E(z,p_2)}\, ,} where $p_1$ and $p_2$ are arbitrary points,
and choose $p_2$ as the point associated to one of the odd spin
structures, say $p_2-\Delta=\delta+\left({0\atop\half}\right)$
\dhok. Now it is straightforward to take a limit similar to the
one that led to \ths. It does not depend on $p_1$, and one finds
$\sigma(z)\propto (z-b)^{-1}(z-\bar a)^{-1}$. The constant factor
is fixed by $SL(2,C)$ covariance, hence \eqn\ssz{\sigma(z)=(z\,
;\bar a, b) \, .}

In Appendix B, we take into account Beltrami differentials dual to
the integration measure \eqn\limm{d\mu(\Omega)=d\tau_1 d\tau_2\,
dl \, ,} and we show that in the degeneration limit, the quantum
determinants appearing in Eq.\mgu\ satisfy:
\eqn\limz{{Z_2\over\widetilde{Z}_2 Z_1^3}={1\over e^{2\pi l}-1}\,
.} This leads to the gaugino mass \eqn\mlead{m_{1/2}=
g^4\int_{F(\Sigma_{\tau})}
     {d\tau_1 d\tau_2\, dl}\, {BZ_{lat\, T^2}\over e^{2\pi l}-1}\, ,}
where we replaced the string coupling by the gauge coupling $g$, with
$g_s=g^2$.

\newsec{Double Degeneration Limit $l\to 0$ on Blown-Up Orbifolds}

It is clear from Eq.\mlead\ that the dominant contribution to the gaugino
mass comes from the region of small $l$.
The $l\to 0$ limit corresponds to the ``double degeneration''
of the surface, with $a\to b$  on \P$^1$. Then the
surface splits into a small neighborhood of $a$ and $b$, of radius $O(l)$,
which comprises the degenerate torus, and the surrounding
annulus extending from the torus to the boundary; in the $l\to 0$
limit the annulus degenerates into a disk with a puncture at $z=a$.
As mentioned before, $l$ cannot
become smaller than  $e^{-\pi\tau_2}$, which is the cutoff provided by the
diameter of the degenerate handle.

The final result for the gaugino mass \mlead\ depends critically
on the small $l$ behavior of the constant $B$. If  $B$ vanishes as
any positive power of $l$, then the integral over $l$ yields a
$\tau_2$-independent constant. On the other hand, if $B$ is
constant, the integral depends logarithmically on the cutoff,
bringing a factor of $\tau_2$ which, as we will see at the end,
will  be converted to $R^2$. Recall that $B$ was introduced in
Eq.\bfad, in the context of $K3$ compactifications, and refers to
a specific basis of two-differentials (C.3) given in Appendix C.

Since the analysis of a general $K3$ turns out to be quite
elaborate, we defer it to another place and return to the orbifold
limit. As mentioned before, the orbifold amplitude \ampAso\
vanishes due to symmetry with respect to the permutations of the
supercurrent insertions. The simplest way to move away from the
orbifold point is by inserting a vertex operator for a massless
``blowing-up'' mode. However, first we explain why the amplitude
is zero for any $\ZZ_N$ orbifold.
\subsec{Vanishing Amplitude on
$\ZZ_N$ orbifolds} After using the explicit parameterization of
the bosonic zero modes, Eqs.(A.1) and (A.5), and summing over all
permutations of the supercurrent insertion points and over all
twisted sectors,\foot{The contribution of the untwisted sector
drops out after symmetrization.} one finds that the amplitude
\ampAso\ contains the factor
\eqn\borb{B_O(z_a)=\sum_{k=1}^{N{-}1}\,\sum_{({ a}_k,{ b}_k)} \,
\sum_{i=1,2} L_{k}L_{{N-k}}L_i\,B_{k, i}(z_a)\, ,}
where $k$ labels
the group element $e^{2\pi ik/N}$, $({a}_k,{b}_k)$ labels the
associated twists along the $\bf a$- and $\bf b$-cycles, and
\eqn\detorbi{B_{k, i}(z_a)=
\det\big[\,\omega_i\,\widetilde{\omega}(z_1)\, ,\,
\omega_{k}\widetilde{\omega}_{{N-k}}(z_2)\, ,\,\omega_{{N-k}}
\widetilde{\omega}_{k}(z_3)\,\big]\, .}
{}First, consider twists that
are non-trivial only along the $\bf b$-cycle, as discussed
explicitly in Appendix A. Then the determinant \detorbi\ is
exactly the same as the determinant computed in Appendix
C. For $\omega_i=\omega_D$ it is zero, while for
$\omega_i=\omega_N$ it is the determinant (C.4), with
a $k$-dependent prefactor ${\rm Im}(h)=\sin(2\pi ik/N)$.
Although Neumann boundary conditions in the SS direction
are beyond the scope of the present
discussion, it is worth mentioning that, even in that case, the final
result would vanish, because
the contribution of the $k$th group element in \borb\ would be
canceled by the $(N-k)$th. For $N$ even, the remaining
contribution of $k=N/2$ vanishes by the same argument as for the
$\ZZ_2$ orbifold mentioned in Section 4. Next,  consider twists
that are non-trivial only along the $\bf a$-cycle. The
corresponding twisted differentials are easy to construct: they
have the property that $\omega_{k}
\widetilde{\omega}_{{N-k}}\propto\omega_{{N-k}}
\widetilde{\omega}_{k}$, therefore the determinant \detorbi\
vanishes also in this case. Similar property holds for twists that
are non-trivial along both $\bf a$- and $\bf b$-cycles. Hence
\eqn\bozero{B_O(z_a)=0\, ,} where the above equation holds to the
same $O(e^{-\pi\tau_2})$ accuracy as the expansions of the
holomorphic differentials.

\subsec{Blowing-up Mode Insertion}

We now insert in the amplitude a blowing-up mode $B$ at zero momentum,
associated to the twisted sector $(h,h^{-1})$ with $h=e^{2i\pi\epsilon}$
and $\epsilon=k/N$. Its vertex in the $-1$ ghost picture is:
\eqn\twistvertex{
V^{(-1,-1)}_B(\zeta,\bar\zeta)=:e^{i\epsilon(\phi_4-\bar\phi_4)
+i(1-\epsilon)(\phi_5-\bar\phi_5)}\sigma_4^{--}\sigma_5^{--}:\, ,
}
where $\sigma_I^{--}$ is the corresponding twist field of conformal
dimension $\epsilon(1-\epsilon)/2$ in both left and right movers. Using
the $N=2$ world-sheet supercurrent, one finds the blowing-up vertex
operator in the 0-ghost picture:
\eqn\twistvertexzero{\eqalign{
V^{(0,0)}_B(\zeta,\bar\zeta)=&:
e^{-i(1-\epsilon)(\phi_4-\phi_5-\bar\phi_4+\bar\phi_5)}
\sigma_4^{++}\sigma_5^{--}+
e^{-i(1-\epsilon)(\phi_4-\phi_5)-i\epsilon(\bar\phi_4-\bar\phi_5)}
\sigma_4^{+-}\sigma_5^{-+}
\cr
&+e^{i\epsilon(\phi_4-\phi_5)+i(1-\epsilon)(\bar\phi_4-\bar\phi_5)}
\sigma_4^{-+}\sigma_5^{+-}+
e^{i\epsilon(\phi_4-\phi_5-\bar\phi_4+\bar\phi_5)}
\sigma_4^{--}\sigma_5^{++}:\, ,
}}
where we used the short distance (OPE) expansions~\dfms:
$\sigma_I^{--}(z,\bar z)\partial
X_I(w)\sim (z-w)^{-1+\epsilon}\sigma_I^{+-}$ and
$\sigma_I^{--}(z,\bar z)\bar\partial X_I^*(w)\sim
(\bar z-\bar w)^{-1+\epsilon}\sigma_I^{-+}$.\foot{The left-right symmetric
twist field $\sigma^{++}$ is called $\tau$ in Ref.~\dfms.} Going from the
$-$ to the $+$ component of the twist field, its conformal dimension is
increased by the corresponding twist ($\epsilon$ or $1-\epsilon$).

After inserting the blowing-up vertex
$\int d^2\zeta V^{(0,0)}_B(\zeta,\bar\zeta)$, the $K3$ part of the
amplitude, corresponding to the last factor of Eq.~\ampA, becomes:
\eqn\Kthreeamp{\eqalign{
&\left\langle e^{i\phi_4/2}(x)e^{i\phi_4/2}(y)e^{-i\phi_4}(z_4)
e^{-i(1-\epsilon)\phi_4}(\zeta)e^{i(1-\epsilon)\bar\phi_4}(\bar\zeta)
\right\rangle_s\, \left\langle
\partial X^4(z_4)\sigma_4^{++}(\zeta,\bar\zeta)\right\rangle\times\cr
&\left\langle e^{i\phi_5/2}(x)e^{i\phi_5/2}(y)e^{-i\phi_5}(z_5)
e^{i(1-\epsilon)\phi_5}(\zeta)e^{-i(1-\epsilon)\bar\phi_5}(\bar\zeta)
\right\rangle_s\, \left\langle
\partial X^5(z_5)\sigma_5^{--}(\zeta,\bar\zeta)\right\rangle\cr
+\, &\left\langle e^{i\phi_4/2}(x)e^{i\phi_4/2}(y)e^{-i\phi_4}(z_4)
e^{i\epsilon\phi_4}(\zeta)e^{-i\epsilon\bar\phi_4}(\bar\zeta)
\right\rangle_s\,
\left\langle\partial X^4(z_4)\sigma_4^{--}(\zeta,\bar\zeta)
\right\rangle\times\cr
&\left\langle e^{i\phi_5/2}(x)e^{i\phi_5/2}(y)e^{-i\phi_5}(z_5)
e^{-i\epsilon\phi_5}(\zeta)e^{i\epsilon\bar\phi_5}(\bar\zeta)
\right\rangle_s\,
\left\langle\partial X^5(z_5)\sigma_5^{++}(\zeta,\bar\zeta)
\right\rangle\cr
\sim\ &\theta_{s,h_4}
\left({1\over 2}(x+y)-z_4-(1-\epsilon)(\zeta-\bar\zeta)\right)
\theta_{s,h_5}
\left({1\over 2}(x+y)-z_5+(1-\epsilon)(\zeta-\bar\zeta)\right)\times\cr
&\left[ {E(z_4,\zeta)E(z_5,\bar\zeta)\over E(z_4,\bar\zeta)E(z_5,\zeta)}
\right]^{1-\epsilon}{1\over E(\zeta,\bar\zeta)^{2(1-\epsilon)^2}}
\left\langle\partial X^4(z_4)\sigma_4^{++}(\zeta,\bar\zeta)\right\rangle
\left\langle\partial X^5(z_5)\sigma_5^{--}(\zeta,\bar\zeta)\right\rangle
\cr +\, &\theta_{s,h_4}
\left({1\over 2}(x+y)-z_4+\epsilon(\zeta-\bar\zeta)\right)
\theta_{s,h_5}
\left({1\over 2}(x+y)-z_5-\epsilon(\zeta-\bar\zeta)\right)\times\cr
&\left[ {E(z_4,\bar\zeta)E(z_5,\zeta)\over E(z_4,\zeta)E(z_5,\bar\zeta)}
\right]^{\epsilon}{1\over E(\zeta,\bar\zeta)^{2\epsilon^2}}
\left\langle\partial X^4(z_4)\sigma_4^{--}(\zeta,\bar\zeta)\right\rangle
\left\langle\partial X^5(z_5)\sigma_5^{++}(\zeta,\bar\zeta)\right\rangle
\, ,}}
where the expression after the $\sim$ symbol replaces the product of the
last two theta-functions in Eq.~\ampAs, together with the
$\prod_{I=4,5}\partial X^I$ factors from the two supercurrent insertions.

The spin structure sum can be performed using the same gauge condition
\gauged\ with the result:
\eqn\Kthreesum{\eqalign{
&\theta_{h_4^{-1}}\left[{0\atop {m\over 2}\vec 1}\right]
\left(z_4+\epsilon(\zeta-\bar\zeta)-\Delta\right)
\theta_{h_5^{-1}}\left[{0\atop {m\over 2}\vec 1}\right]
\left(z_5-\epsilon(\zeta-\bar\zeta)-\Delta\right)\cr
&\left[ {E(z_4,\bar\zeta)E(z_5,\zeta)\over E(z_4,\zeta)E(z_5,\bar\zeta)}
\right]^{\epsilon}{1\over E(\zeta,\bar\zeta)^{2\epsilon^2}}
\left\langle\partial X^4(z_4)\sigma_4^{--}(\zeta,\bar\zeta)
\right\rangle_{h_4}
\left\langle\partial X^5(z_5)\sigma_5^{++}(\zeta,\bar\zeta)
\right\rangle_{h_5}\cr
+\ &(4\leftrightarrow 5, \epsilon\leftrightarrow 1-\epsilon)
\, ,}}
which has to replace the last two terms in the product of Eq.~\ampAso.

We want to show that after integrating Eq.\Kthreesum\
over the vertex position $\zeta$, the result is {\it not}
symmetric under the transposition $z_4\leftrightarrow z_5$; this will be true
even in the untwisted sector with $h_4=h_5=1$. To that end, we can go
directly to the double degeneration limit $l\to 0$. Recall that the vertex
position is to be integrated
over the whole half-plane, ${\rm Im}\zeta> 0$. If one is interested
only in terms that are non-vanishing in the
$l\to 0$ limit, one can neglect the small [radius $O(l)$] neighborhood of
the puncture points.
This means that the contribution of the factorized torus is negligible and
that the vertex is {\it de facto} inserted on a hemi-sphere
with one puncture point $a$. Then one can set
$\zeta= 0$ in the arguments of theta functions. Furthermore, the twist
field correlators should be evaluated on such a hemisphere. Due to the
$\ZZ_N$ symmetry of the lattice, only the ``quantum'' parts
of these correlators \atick\ contribute to the amplitude.\foot{Actually,
the vanishing due to the $\ZZ_N$ symmetry can be avoided by turning on a
non-diagonal modulus mixing the two planes. However, the resulting
contribution would be exponentially suppressed in the $\tau_2\to\infty$
limit.} Their form is completely determined by OPE and $SL(2,R)$
invariance~\refs{\dfms,\atick}. Without going into much detail, it is
clear that Eq.\Kthreesum\ is not symmetric in $z_4,z_5$: in the last line,
$z_4\leftrightarrow z_5$ is accompanied by
$\epsilon\leftrightarrow 1-\epsilon$, and this
asymmetry should survive the $\zeta$-integration. With the
twist correlators non-vanishing on a punctured hemisphere,
the constant $B(l=0)$ will be non-zero on blown-up orbifolds.

\newsec{Radius Dependence of the Gaugino Mass}

We return to the gaugino mass \mlead, now turning our attention to the
dependence of the amplitude on the string
configuration on $T^2$. Eq.\bfac\ contains the respective  zero mode,
$\partial X_3=\partial X+i\partial X'$,
where $X$ is the coordinate in the SS direction and $X'$ is the
``non-SS'' coordinate on $T^2$. For the zero-mode configurations
parameterized as in Eq.(A.1),
\eqn\dX{
\partial X(z)=2\pi R\sum_{i=1,2}L_i\,\omega_i(z)\, ,}
where the constants $L_i$ depend on the winding numbers $m$ and $n$,
see Eq.(A.3).
The SS coordinate satisfies D boundary conditions and, as explained
before, in the $\tau_2\to\infty$ limit all $n\neq 0$ modes are
exponentially suppressed by the partition function, hence
\eqn\dXa{\partial X={m\pi R\over i(\tau_2-
l)}\,\omega_D(z)~,\qquad\omega_D(z)=\omega_1(z)+\omega_2(z).}
Furthermore, non-trivial $\partial X'$ modes are also exponentially
suppressed in this limit,
unless the corresponding radius blows up in the same rate as $R$. Since
the final result is the same, we will not consider here such a
``fine-tuned" scenario. In this way, we obtain
\eqn\Bas{
B\propto m{R\over\tau_2 -l}}
where we ignored the factors associated to $K3$ which had already
been discussed in the previous section.

The zero modes are weighted by the torus partition function.
After combining Eq.\Bas\ with \zerstlim\ and \Zb, we obtain
\eqn\Zlat{
BZ_{lat\, T^2}\propto { R^2\over (\tau_2-l)^{3/2}}{m\over (\tau_2+l)^2}\,
  \exp({\displaystyle -{m^2\pi R^2\over \tau_2-l}})\, .}
It is clear that the amplitude vanishes after summing over the
windings, due to the $\ZZ_2$ symmetry $m\to -m$. Since this
symmetry is absent in the presence of Wilson lines, one could try
to insert into the amplitude \ampAM\ a zero-momentum vertex for
the Wilson line modulus $\phi$:
\eqn\wilsa{\eqalign{\widetilde\omega(x)\,\widetilde\omega(y)
\longrightarrow &\,
\widetilde\omega(x)\big[\phi\int_y^xdu\,\partial X_3(u)\big]\,
\widetilde\omega(y)\cr \simeq &\,
\widetilde\omega(x)\widetilde\omega(y)\int_y^xdu\,\omega_D(u)\cr
\simeq &\, \widetilde\omega(x)\widetilde\omega(y)\left[
\Omega_D(x)-\Omega_D(y)\right]\, ,}} where in the second line we
replaced $\partial X_3(u)$ by its zero mode \dXa, since it cannot
be contracted with other operators, while in the third line we
used the property that $\omega_D(u)$ is a closed one-form to write
$\omega_D(u)\equiv d\Omega_D(u)$. The above result obviously
vanishes again upon integration over $x$ and $y$.

\subsec{Generalized Scherk-Schwarz Deformation}

To avoid the vanishing of the summation over windings of the SS
circle, we consider a $\ZZ_N$ SS deformation which is a
straightforward generalization of the $\ZZ_2$ case~\kpor. For
instance, consider compactifications on the product space of a
$K3$ orbifold times $T^2$, and define the internal rotation
current $J$, in one of the two complex coordinates of $K3$:
\eqn\current{
J=\psi_4^*\psi_4+X_4^*{\buildrel\leftrightarrow\over\partial}
X_4\, .} Obviously, only discrete rotations remain that correspond
to symmetries of the compactification lattice, generated by
$Q=\exp\{2i\pi e\oint J\}$ with $e=1/N$ for a $\ZZ_N$ symmetry. A
generalized SS deformation can then me obtained by coupling the
rotation charges with the winding numbers of a circle from $T^2$
(or with the whole $T^2$ momentum lattice in general). The
corresponding partition function is obtained by replacing:
\eqn\defssgen{ {\Theta\left[{{\vec a}\atop{\vec b}}\right]\over
Z_1\left[{h_{\bf a}\atop h_{\bf b}}\right]}\longrightarrow
e^{-2i\pi en(b_1-b_2)}\ {\Theta\left[{{\vec a}+ne{\vec
1}\atop{\vec b}+me{\vec 1}}\right]\over Z_1\left[{h_{\rm\bf
a}+ne{\vec 1}\atop h_{\rm\bf b}+me{\vec 1}}\right]} \, , } where
$Z_1$ denotes the contribution of the $(h_{\rm\bf a},h_{\rm\bf
b})$ twisted sector. Eq.~\defssgen\ generalizes \ssdef, which is
recovered by setting $e=1$, corresponding to a $2\pi$ rotation. In
this case, the bosonic part is not modified and the deformation
amounts to inserting the phase $\delta^{\rm SS}_s$.

Following the reasoning of Section 3, in the large radius limit,
only vanishing windings $n=0$ may give corrections that escape
exponential suppression in the degeneration limit
$\tau_2\to\infty$. Thus, only the boundary conditions along the
$\bf b$ cycle are modified in \defssgen, and the spin
structure-dependent part of the amplitude \ssorbi\ (with the
picture-changing insertion points satisfying the same gauge
condition \gauged) becomes: \eqn\ssorbig{\eqalign{{\cal
S_O}\longrightarrow \sum_{\vec a,\vec b} &~~~ \theta\left[{{\vec
a}\atop{\vec b}}\right]({1\over 2}(x-y))\, \theta\left[{{\vec
a}\atop{\vec b}}\right]({1\over 2}(x+y)-z_3)\cr
&\times\theta_{h}\left[{{\vec a}\atop{\vec b}+me{\vec 1}}\right]
({1\over 2}(x+y)-z_4)\, \theta_{h^{-1}}\left[{{\vec a}\atop{\vec
b}}\right]({1\over 2}(x+y)-z_5) \, ,}} where we defined, as in
section 4.1, $h_4\equiv h$ and thus $h_3=1$ and $h_5=h^{-1}$ for
an internal space $T^2\times T^4/H$. The spin structure sum can be
performed using the Riemann identity with the result: \eqn\ssorbjg
{{\cal S_O}\longrightarrow \theta\left[{0\atop {m\over 2}e\vec
1}\right](x-\Delta)\, \theta\left[{0\atop {m\over 2}e\vec
1}\right](z_3-\Delta)\, \theta_{h^{-1}}\left[{0\atop -{m\over
2}e\vec 1}\right](z_4-\Delta)\, \theta_h\left[{0\atop {m\over
2}e\vec 1}\right](z_5-\Delta) \, .} After incorporating the
remaining factors, one finds that the amplitude \ampAso\ becomes:
\eqn\ampAsog{\eqalign{ {\cal A}\longrightarrow
~&\theta\left[{0\atop {m\over 2}e\vec 1}\right](x-\Delta)
{\sigma(x)\sigma(y)\over\prod_{I<J}^{3,4,5}E(z_I,z_J)
\prod_{I=3}^5\sigma^2(z_I)}\times {Z_2\over Z_1^5Z_{1,h\oplus
me}Z_{1,h^{-1}}}\cr \times &\, \theta\left[{0\atop {m\over 2}e\vec
1}\right](z_3-\Delta) \theta_{h^{-1}}\left[{0\atop -{m\over
2}e\vec 1}\right](z_4-\Delta)\, \theta_h\left[{0\atop {m\over
2}e\vec 1}\right](z_5-\Delta)\cr \times &\, \partial X^3
(z_3)\,\partial X^4_{h\oplus me}(z_4)\,
\partial X^5_{h^{-1}}(z_5)\times Z_{lat}\, ,}}
where the symbol $h\oplus me$ denotes the addition of $me$ to the
twist $h\equiv (h_{\rm\bf a},h_{\rm\bf b})$ along the ${\bf b}$
cycle only: $h\oplus me\equiv (h_{\bf a}, h_{\bf b}e^{2i\pi me})$.
Note that the $\bf b$-cycle periodicity conditions for all
$z_I$'s, {\it including} $z_4$, and $x$ are twisted by the same
$\ZZ_{2N}$ group element $e^{i\pi ke}=e^{i\pi k/N}$ for $m=k~
mod~2N$. The contribution of the $k=0$ class of windings
disappears by Riemann vanishing theorem (supersymmetry), while the
class of $k=N$ has the same structure as odd windings in the
standard SS mechanism, and disappears after summation as a
consequence of $m\to -m$ symmetry. Since the zero modes of
$\partial X^3$ are linear in these winding numbers, the remaining
contributions of $m=k~ mod~2N$ come with opposite sign to those of
$m=2N-k~ mod~2N$; however, as explained below, they are weighted by
different factors.

Starting from the amplitude \ampAsog\ and repeating the same
steps that led us in Section 4 to the final expression \ampAM:
antisymmetrization, bosonisation formula, {\it etc.}, for each
class of $k~ mod~ 2N$ windings, one obtains a similar expression,
but now with the tilde representing a twist by $e^{i\pi k/N}$
along the $\bf b$ cycle, {\it i.e.}
\eqn\newo{\widetilde{\omega}(x)\longrightarrow\omega_k(x)~,\qquad
\widetilde{\omega}(y)\longrightarrow\omega_k(y)\, .} The
integration over the vertex positions yields
\eqn\omtw{\left(\int_{\bf c}\omega_k(x)\right)^2=
(1-e^{i\pi k/N})^2.}
Note that the class of $2N-k~ mod~ 2N$ windings (which
come with the opposite sign) is weighted by the complex conjugate
of the above factor. Now it is clear that the sum over the SS
winding numbers does not vanish. However, one still needs to blow
up singularities as in Section 6, in order to avoid cancellations
due to similar symmetries of the $K3$ orbifold sector.

\subsec{Determination of the Gaugino Mass}

At this point, we can exclude any ``accidental'' cancellation and
return to Eq.\Zlat, assuming that the sum over $m$ is restricted
to one of the $m=k ~mod~ 2N$ ($k=1,\dots,N-1$) class of
windings. Then Eq.\mlead\ gives \eqn\mgz{m_{1/2}\propto g^4\int
d\tau_2\int { dl\over (e^{2\pi l}-1)} {1\over
(\tau_2-l)^{3/2}}\sum_{m} {mR^2\over (\tau_2+l)^2}\,
\exp({\displaystyle -{m^2\pi R^2\over \tau_2-l}})\, . } The
integral over $l$ has a logarithmic divergence at $l=0$ which, as
explained in Section 6, is regulated by $e^{-\pi\tau_2}$. This
brings a (large) factor of $\tau_2$, while $l$ can be set zero
everywhere else in the integrand, so that
\eqn\mgy{m_{1/2}\propto
g^4\int {d\tau_2\over \tau_2^{5/2}}\sum_{m}
mR^2\exp({\displaystyle -{m^2\pi R^2\over \tau_2}})\, .}
After rescaling $\tau_2\to \tau_2 R^2$, the above expression yields
\eqn\mgw{m_{1/2} \;\propto\; {g^4\over R} \;\propto\;
g^4m_{3/2}\, .}

The mass \mgw\ can be understood within the effective field theory
by looking at a generic one-loop graph involving a gravitational
exchange. Each vertex brings one power of the Plank mass
in the denominator, and thus
$m_{1/2}\simeq m_{3/2}^3/M_P^2$ if the loop momentum integral is
convergent~\aq. However if, as na\"{\i}vely expected, the
momentum integral is quadratically divergent, then one obtains
$m_{1/2}\simeq m_{3/2}\,\Lambda^2_{UV}/M_P^2$. Since the
ultraviolet cutoff $\Lambda_{UV}\propto M_P$, the result \mgw\
confirms this expectation. Note, however, that in the case of
orbifold compactifications, for which $m_{1/2}={\cal
O}(e^{-1/\alpha' m_{3/2}^2})$, such a crude argument fails because it
ignores the effects of discrete symmetries.

\newsec{Conclusions and Outlook}

In this work, we developed a formalism for
computing one-loop gravitational corrections to the effective
action of D-branes. Furthermore, we
have shown that the genus 3/2 amplitude
responsible for communicating the bulk supersymmetry breaking to
open string fermions is closely related to the well-known genus 2
topological term ${\cal F}_2$. For models with large extra
dimensions and low-energy supersymmetry breaking \`a la
Scherk-Schwarz, this mass is proportional to the gravitino mass.
However, the result is zero in the case of the standard,
$\ZZ_2$-symmetric, Scherk-Schwarz mechanism, hence one is forced
to consider its $\ZZ_N$ generalization in order to generate a
non-vanishing gaugino mass.

In fact, it is not easy to find explicit examples of such a mass
generation in ``calculable'' orbifold models. Orbifolds have
discrete symmetries that generically lead to further
cancellations. These, in turn, can be avoided by blowing up the
orbifold singularities, which can be accomplished  by switching on
non-zero VEVs of certain twisted fields. This mass generation
mechanism clearly needs a deeper explanation. On one hand, one
should find a more profound connection to the underlying
world-sheet theory. On the other hand, it would be very useful to
have a more precise field-theoretical description in terms of
effective interactions and Feynman diagrams. Both the topological
origin as well as the relation to quadratic divergences strongly
suggest that the mass is generated by anomalies, but  more work is
needed to identify a direct link.

One could think that the gaugino masses are due to the
so-called ``anomaly-mediation'' \rs\ of the four-dimensional
(super)conformal anomaly. However, such a mechanism predicts
one-loop masses proportional to $g^2$ and not to $g^4$ as we find. Furthermore,
 the anomaly-mediated gaugino masses are
proportional to the beta functions, therefore they would require
the presence of at least one more world-sheet boundary.
Thus the mismatch of the powers of gauge coupling and of the group-theoretical
coefficients hints against superconformal anomaly mediation.
In fact, our results
raise an interesting question whether the anomaly mediation
mechanism can be implemented at all in any theory with a sensible
ultraviolet completion.

\vskip 1cm
\centerline{\bf Acknowledgments} \vskip 3mm\noindent
This work would have never been completed without the help of
Narain, who had assisted us at its crucial stages. We are deeply indebted
to him. We are grateful to Pierre Vanhove and Frank Ferrari for their
help in resolving several technical problems. We acknowledge illuminating
discussions with Louis Alvarez-Gaum\'e, Carlo Angelantonj, Edi Gava,
Stephan Stieberger and Cumrun Vafa. T.R.T.\ is grateful to CERN Theory Division
for its hospitality during the sabbatical leave of 2003, when most of this work
was done. His work  is supported in part by the National Science
Foundation under grant PHY-02-42834. The work of I.A.\ is
supported in part by the European Commission under the RTN
contract HPRN-CT-2000-00148.

\appendix{A}{Bosonic Zero Modes and Twisted Differentials}

The amplitude that determines the gaugino mass includes
three supercurrent insertions, with one of them effectively coupled to the
the SS torus $T^2$ and the other two to the $K3$ surface.
The $T^2$ contribution involves an explicit factor of the (untwisted)
zero mode $\partial X_3$; in the case
of orbifold compactifications, the other two insertions involve also the
zero modes associated to the twisted sectors, see respectively
Eqs.\ampAsK\ and \ampAs.

In the case of untwisted direction, it is convenient to consider
separately the two real components, each of the form
\eqn\ssx{X(P)=2\pi R\sum_{i=1,2}[L_i\int^P_O\omega_i(z) +\bar
L_i\int^P_O\bar\omega_i(\bar z)]\, ,}
where $O$ is an arbitrary base point. Due to the reality property
\reality, the Neumann and Dirichlet boundary conditions read,
respectively,
\eqn\lpro{{\bar L}_{1,2}=\pm L_{2,1}\qquad \left\{
{+~{\rm for}~N\atop -~{\rm for}~D}\right.}
By considering the
boundary conditions for a string winding $n$ times around the $\bf
a$-cycle and $m$ times around the $\bf b$-cycle, we obtain
\eqn\luntw{L_1={m+n[\tau_1-i(\tau_2\pm l)]\over 2i(\tau_2\pm l)}\,
,\qquad L_2=\pm {\bar L}_1\, .} The classical action, $S_{\rm
cl}={1\over 2\pi\alpha'}\int_{\Sigma_1}\partial_zX\partial_{\bar
z}X$, can be
     computed by using Eqs.\integrals, with the result
written in Eq.\Sclass. As explained in Section 3, in the
degeneration limit $\tau_2\to \infty$, $l$ fixed, the $n\neq 0$
modes are exponentially suppressed by the partition function,
therefore the relevant zero modes are given by
\eqn\dxmodes{\partial_z X_{N,D}={m\pi R\over i(\tau_2\pm
l)}\omega_{N,D}\, ,\qquad\quad \omega_{N,D}=\omega_1(z)\mp\omega_2(z)\, .}

In order to construct the zero modes twisted by a $\ZZ_N$ group element
$h$, we can start from the $g=2$ double cover $\Sigma$, with the twisted
{\it complex} instanton configuration \narain
\eqn\ssxtw{X_h(P)=2\pi R[L_h\int^P_O\omega_h(z)
+\bar{L}_h\int^P_O\bar\omega_{h^*}(\bar z)]\, ,}
where
$\omega_h(z)$ is the (unique) holomorphic one-differential twisted by
$(h_{a_i}, h_{b_i}), ~i=1,2$, along the $\bf a$- and $\bf b$-cycles of
$\Sigma$, respectively. The constants $L_h$ and $\bar{L}_h$ are
determined by the boundary conditions. Since we are interested in the
instanton solutions on $\Sigma_1$,  we identify $h_{a_1}=h_{\bf a}$,
$h_{b_1}=h_{\bf b}$. For our purposes, it is completely sufficient
to consider configurations with a
non-trivial twist only along the $\bf b$-cycle: $h_{\bf a}=1,~ h_{\bf
b}\equiv h$.

The simplest way to obtain the $\tau_2\to\infty$ limits
of twisted differentials is
by working directly in the degeneration limit of the punctured hemisphere.
The (unique) $SL(2,C)$-covariant, holomorphic differential twisted
by $h$ along the $\bf b$-cycle is given by:
\eqn\sigtw{\omega_h(z)=(z;a,\bar a)-h\, (z;b,\bar b)\, ,}
where $(z;p_1,p_2)$ is defined in \hdif.
If an additional twist by $(-1)$ along the $\bf b$-cycle is present, then
\eqn\sigtwb{\widetilde\omega_h(z)=(z;a,\bar a)+h\, (z;b,\bar b)\, .}
It is easy to see that
$\widetilde\omega(z)=\widetilde\omega_{h=1}(z)$ given by the above
formula does indeed
appear in the $\tau_2\to\infty$ limit of Eq.\omtil.
This also proves Eq.\limo. Note that at the boundary,
\eqn\parit{\omega_h(x)={\bar \omega}_{h^*}(x)\, ,\qquad x\in
\partial\Sigma_1,}
therefore the Neumann and Dirichlet boundary conditions read, respectively,
\eqn\dntw{\bar{L}_h=\pm L_h\qquad \left\{
{+~{\rm for}~N\atop -~{\rm for}~D}\right.\, .}
Since the amplitude considered in the paper vanishes for orbifold
compactifications, we do not pursue the discussion of twisted
instanton configurations any further. We will make use, however,
of the twisted differentials in the following
computation of $\widetilde{Z}_2$.

\appendix{B}{Moduli Integration Measure}
The integration measure over the moduli of the $(g=1,h=1)$
Riemann surface under consideration is induced by the chiral
measure of its $g=2$ double-cover. It involves 3 insertions of
the reparametrization ghosts $b$ necessary to soak up the
corresponding zero modes: \eqn\meas{d\mu(\Omega)=d\tau d\bar\tau
dl\;Z_2^{-1}\epsilon^{a_1a_2a_3} \int\left<\prod_{p=1}^{p=3} dw_p\,
\mu_{a_p}(w_p)b(w_p)\right>,} where $\mu_{a_p}$, $p=1,2,3$, are
the Beltrami differentials dual to the moduli
$(d\tau=d\Omega_{11}\, ,d\bar\tau=d\Omega_{22}\, ,dl=
d\Omega_{12})$, respectively.  The correlation function on the
r.h.s., \eqn\three{\left\langle
b(w_1)b(w_2)b(w_3)\right\rangle\equiv Z_2(w_1,w_2,w_3)
=Z_2\det\omega_i\omega_j(w_p)\, ,} where the constant factor $Z_2$
can be considered as the (oscillator part of the) partition
function of the (2,$-$1) $b{-}c$ system. This determinant has already been
included in the amplitude \ampAs, therefore in order to avoid double counting
we inserted the factor $Z_2^{-1}$ in \meas.
The three quadratic
differentials, $\omega_1\omega_1$, $\omega_1\omega_2$ and
$\omega_2\omega_2$ form a basis of the $b$-ghost zero modes. After
inserting \three\ into the r.h.s.\ of \meas, the position
integration can be performed explicitly; since the Beltrami
differentials are dual to $d\Omega_{ij}$,
\eqn\mes{d\mu(\Omega)=\, d\tau_1 d\tau_2\, dl\, .}

In order to determine the large $\tau_2$ behavior of $Z_2$, it is
convenient to use the bosonisation formula \ver:
\eqn\zcor{Z_2(w_1,w_2,w_3)=Z_1^{-1/2}\theta(\sum_{p=1}^{p=3}w_p-3\Delta)
\prod_{p<q}^{1,2,3}E(w_p,w_q)\prod_{p=1}^{p=3}\sigma^3(w_p)\, ,} where the
factor $Z_1^{-1/2}$ is the (oscillator part of the) partition
function of a chiral scalar field. The argument of the theta
function can be rewritten as:
\eqn\rewr{\sum_{p=1}^{p=3}w_p-3\Delta=\!\vec{\,\xi}+3\delta~, \qquad\qquad
\!\vec{\,\xi}=\sum_{p=1}^{p=3}w_p-3\Delta_{\delta}~.} Since $\!\vec{\,\xi}$
is finite in
the $\tau_2\to\infty$ limit, the leading term of the theta series
is \eqn\lead{\theta(\sum_{p=1}^{p=3}w_p-3\Delta)=e^{4\pi
(\tau_2-l)}f(\!\vec{\,\xi})
\, ,} where
\eqn\zved{ f(\!\vec{\,\xi})=
e^{-2\pi(\xi_1+\xi_2)}[1+e^{-\pi l}e^{-2\pi i\xi_1}+e^{-\pi
l}e^{-2\pi i\xi_2}+e^{-2\pi i(\xi_1+\xi_2)}]\,  .}

It is a matter of
straightforward but quite tedious algebra to show that for the differentials
\hdif,
\eqn\middet{\det\omega_i\omega_j(w_p)=f(\!\vec{\,\xi})\,
\,(e^{2\pi l}-1)^2\, e^{-6\pi l}
\prod_{p<q}^{1,2,3} E(w_p,w_q)\prod_{p=1}^{p=3}\sigma^3(w_p)\, .}
Hence, after comparing Eqs.\three\ and
\zcor, we obtain \eqn\zzb{Z_2=Z_1^{-1/2}\,{e^{4\pi
(\tau_2-l)}e^{6\pi l} \over(e^{2\pi l}-1)^2}\, .}

At this point, it remains to determine the large $\tau_2$ limit of
the chiral boson partition function $Z_1$. To that end, it is
convenient to use the chiral bosonisation formula \ver:
\eqn\chb{Z_1^{3/2}={\theta(u_1+u_2-v-\Delta)
\over\det\omega_i(u_j)}{E(u_1,u_2)\over
E(u_1,v)E(u_2,v)}{\sigma(u_1)\sigma(u_2)\over\sigma(v)}\, ,} where
$u_1$, $u_2$ and $v$ are arbitrary. After taking the limit in a
similar way as before, we obtain \eqn\zzx{Z_1=1\, .}

In order to compare with the existing literature,
we note that, according to our results, the
``two-loop cosmological constant" of $D=26$ bosonic string,
\eqn\muo{{Z_2\over Z^{13}_1}\, d\mu(\Omega)=d\tau_1 d\tau_2\, dl\,{e^{4\pi
(\tau_2-l)} e^{6\pi l} \over(e^{2\pi l}-1)^2}\, ,}
where we used Eqs.\mes, \zzb\ and \zzx.
This quantity should be compared with the formal square root
of the well-known  result
\refs{\morozov,\moore}: \eqn\muol{{Z_2\over Z^{13}_1}\,d\mu(\Omega)=\prod_{i\le
j}{d\Omega_{ij}\over |\Psi_{10}(\Omega)|}\, ,} where $\Psi_{10}$
is the weight 10 generator of the Isgusa ring,
$\Psi_{10}(\Omega)=\prod_{\xi}\theta^2[\xi](\Omega, 0)$, with the
product including all 10 even spin structures. Indeed, Eq.\muo\
follows in the $\tau_2\to\infty$ ($l$ fixed)  degeneration limit of
$\Psi_{10}$. The double pole appearing in the factorization limit
$\Omega_{12}=l=0$ can be attributed to the tachyon \moore.

\appendix{C}{Twisted Determinant $\bf \widetilde{Z}_2$}
In order to find the $\tau_2\to \infty$ limit
of the  determinant $\widetilde{Z}_2$, we will compare the
degeneration limits of the left- and right-hand sides of Eq.\bosver.

On the l.h.s., we have a theta function series, with the leading term
\eqn\thlim{\theta
\left[{0\atop {1\over 2}\vec 1}\right](\sum_{a=1}^{a=3} z_a-3\Delta)
=e^{4\pi(\tau_2-l)}\tilde f(\!\vec{\,\zeta})~,\qquad\quad  \!\vec{\,\zeta}=
\sum_{a=1}^{a=3} z_a-3\Delta_\delta~,}
where
\eqn\zvec{\tilde f(\!\vec{\,\zeta})=
e^{-2\pi(\zeta_1+\zeta_2)}[1-e^{-\pi l}e^{-2\pi i\zeta_1}-e^{-\pi
l}e^{-2\pi i\zeta_2}+e^{-2\pi i(\zeta_1+\zeta_2)}]\, .}
Furthermore, the
degeneration limits of the prime form and the $\sigma$-differential are
given by Eqs.\prif\ and \ssz, respectively, and $Z_1=1$,
as shown in Appendix B.

On the r.h.s., we have the determinant $\det\widetilde{h}_a(z_b)$ involving
twisted two-differentials $\widetilde{h}_a$, $a=1,2,3$. We can compute this
determinant explicitly by choosing the basis
\eqn\hbase{\widetilde h_1=\omega_N\,\widetilde\omega\qquad \widetilde
h_2=\omega_h\,
\widetilde\omega_{h^{-1}}\qquad \widetilde h_3=\omega_{h^{-1}}
\widetilde\omega_{h}\, ,}
with the $\omega$-differentials listed in Eqs.\dxmodes,
\sigtw, \sigtwb\ and \hdif.
After a straightforward but tedious computation, we find that, up to
a numerical
factor,
\eqn\findet{\det\widetilde{h}_a(z_b)={\rm Im}(h)\times
\,\tilde f(\!\vec{\,\zeta})\,(e^{2\pi l}-1)\, e^{-6\pi l}
\prod_{a<b}^{1,2,3} E(z_a,z_b)\prod_{a=1}^{a=3}\sigma^3(z_a)\, .}

By comparing the two sides of Eq.\bosver, we find:
\eqn\twistd{\widetilde{Z}_2={e^{4\pi(\tau_2-l)}\, e^{6\pi l}\over
e^{2\pi l}-1}\, .} Then Eq.\limz\ follows after combining the
above result with Eqs.\zzb\ and \zzx. The ${\rm Im}(h)$ factor,
which reflects antisymmetry of the determinant under
$h\leftrightarrow h^{-1}$ for a specific choice of basis (C.3), is
irrelevant to the present derivation, however it can be important
when summing over twisted sectors, as pointed out in Section 6.

\vfill\eject

     \listrefs

\bye
\end